# The IceCube Neutrino Observatory Part V: Neutrino Oscillations and Supernova Searches

THE ICECUBE COLLABORATION

## Contents

# IceCube Collaboration Member List

M. G. Aartsen[2], R. Abbasi[27], Y. Abdou[22], M. Ackermann[41], J. Adams[15], J. A. Aguilar[21], M. Ahlers[27], D. Altmann[9], J. Auffenberg[27], X. Bai[31,a], M. Baker[27], S. W. Barwick[23], V. Baum[28], R. Bay[7], J. J. Beatty[17,18], S. Bechet[12], J. Becker Tjus[10], K.-H. Becker[40], M. Bell[38], M. L. Benabderrahmane[41], S. BenZvi[27], P. Berghaus[41], D. Berley[16], E. Bernardini[41], A. Bernhard[30], D. Bertrand[12], D. Z. Besson[25], G. Binder[8,7], D. Bindig[40], M. Bissok[1], E. Blaufuss[16], J. Blumenthal[1], D. J. Boersma[39], S. Bohaichuk[20], C. Bohm[34], D. Bose[13], S. Böser[11], O. Botner[39], L. Brayeur[13], H.-P. Bretz[41], A. M. Brown[15], R. Bruijn[24], J. Brunner[41], M. Carson[22], J. Casey[5], M. Casier[13], D. Chirkin[27], A. Christov[21], B. Christy[16], K. Clark[38], F. Clevermann[19], S. Coenders[1], S. Cohen[24], D. F. Cowen[38,37], A. H. Cruz Silva[41], M. Danninger[34], J. Daughhetee[5], J. C. Davis[17], C. De Clercq[13], S. De Ridder[22], P. Desiati[27], K. D. de Vries[13], M. de With[9], T. DeYoung[38], J. C. Díaz-Vélez[27], M. Dunkman[38], R. Eagan[38], B. Eberhardt[28], J. Eisch[27], R. W. Ellsworth[16], S. Euler[1], P. A. Evenson[31], O. Fadiran[27], A. R. Fazely[6], A. Fedynitch[10], J. Feintzeig[27], T. Feusels[22], K. Filimonov[7], C. Finley[34], T. Fischer-Wasels[40], S. Flis[34], A. Franckowiak[11], K. Frantzen[19], T. Fuchs[19], T. K. Gaisser[31], J. Gallagher[26], L. Gerhardt[8,7], L. Gladstone[27], T. Glüsenkamp[41], A. Goldschmidt[8], G. Golup[13], J. G. Gonzalez[31], J. A. Goodman[16], D. Góra[41], D. T. Grandmont[20], D. Grant[20], A. Groß[30], C. Ha[8,7], A. Haj Ismail[22], P. Hallen[1], A. Hallgren[39], F. Halzen[27], K. Hanson[12], D. Heereman[12], D. Heinen[1], K. Helbing[40], R. Hellauer[16], S. Hickford[15], G. C. Hill[2], K. D. Hoffman[16], R. Hoffmann[40], A. Homeier[11], K. Hoshina[27], W. Huelsnitz[16,b], P. O. Hulth[34], K. Hultqvist[34], S. Hussain[31], A. Ishihara[14], E. Jacobi[41], J. Jacobsen[27], K. Jagielski[1], G. S. Japaridze[4], K. Jero[27], O. Jlelati[22], B. Kaminsky[41], A. Kappes[9], T. Karg[41], A. Karle[27], J. L. Kelley[27], J. Kiryluk[35], J. Kläs[40], S. R. Klein[8,7], J.-H. Köhne[19], G. Kohnen[29], H. Kolanoski[9], L. Köpke[28], C. Kopper[27], S. Kopper[40], D. J. Koskinen[38], M. Kowalski[11], M. Krasberg[27], K. Krings[1], G. Kroll[28], J. Kunnen[13], N. Kurahashi[27], T. Kuwabara[31], M. Labare[22], H. Landsman[27], M. J. Larson[36], M. Lesiak-Bzdak[35], M. Leuermann[1], J. Leute[30], J. Lünemann[28], J. Madsen[33], G. Maggi[13], R. Maruyama[27], K. Mase[14], H. S. Matis[8], F. McNally[27], K. Meagher[16], M. Merck[27], P. Mészáros[37,38], T. Meures[12], S. Miarecki[8,7], E. Middell[41], N. Milke[19], J. Miller[13], L. Mohrmann[41], T. Montaruli[21,c], R. Morse[27], R. Nahnhauer[41], U. Naumann[40], H. Niederhausen[35], S. C. Nowicki[20], D. R. Nygren[8], A. Obertacke[40], S. Odrowski[20], A. Olivas[16], M. Olivo[10], A. O'Murchadha[12], L. Paul[1], J. A. Pepper[36], C. Pérez de los Heros[39], C. Pfendner[17], D. Pieloth[19], E. Pinat[12], J. Posselt[40], P. B. Price[7], G. T. Przybylski[8], L. Rädel[1], M. Rameez[21], K. Rawlins[3], P. Redl[16], R. Reimann[1], E. Resconi[30], W. Rhode[19], M. Ribordy[24], M. Richman[16], B. Riedel[27], J. P. Rodrigues[27], C. Rott[17,d], T. Ruhe[19], B. Ruzybayev[31], D. Ryckbosch[22], S. M. Saba[10], T. Salameh[38], H.-G. Sander[28], M. Santander[27], S. Sarkar[32], K. Schatto[28], M. Scheel[1], F. Scheriau[19], T. Schmidt[16], M. Schmitz[19], S. Schoenen[1], S. Schöneberg[10], A. Schönwald[41], A. Schukraft[1], L. Schulte[11], O. Schulz[30], D. Seckel[31], Y. Sestayo[30], S. Seunarine[33], R. Shanidze[41], C. Sheremata[20], M. W. E. Smith[38], D. Soldin[40], G. M. Spiczak[33], C. Spiering[41], M. Stamatikos[17,e], T. Stanev[31], A. Stasik[11], T. Stezelberger[8], R. G. Stokstad[8], A. Stößl[41], E. A. Strahler[13], R. Ström[39], G. W. Sullivan[16], H. Taavola[39], I. Taboada[5], A. Tamburro[31], A. Tepe[40], S. Ter-Antonyan[6], G. Tešić[38], S. Tilav[31], P. A. Toale[36], S. Toscano[27], M. Usner[11], D. van der Drift[8,7], N. van Eijndhoven[13], A. Van Overloop[22], J. van Santen[27], M. Vehring[1], M. Voge[11], M. Vraeghe[22], C. Walck[34], T. Waldenmaier[9], M. Wallraff[1], R. Wasserman[38], Ch. Weaver[27], M. Wellons[27], C. Wendt[27], S. Westerhoff[27], N. Whitehorn[27], K. Wiebe[28], C. H. Wiebusch[1], D. R. Williams[36], H. Wissing[16], M. Wolf[34], T. R. Wood[20], K. Woschnagg[7], D. L. Xu[36], X. W. Xu[6], J. P. Yanez[41], G. Yodh[23], S. Yoshida[14], P. Zarzhitsky[36], J. Ziemann[19], S. Zierke[1], M. Zoll[34]

[1]III. Physikalisches Institut, RWTH Aachen University, D-52056 Aachen, Germany
[2]School of Chemistry & Physics, University of Adelaide, Adelaide SA, 5005 Australia
[3]Dept. of Physics and Astronomy, University of Alaska Anchorage, 3211 Providence Dr., Anchorage, AK 99508, USA
[4]CTSPS, Clark-Atlanta University, Atlanta, GA 30314, USA
[5]School of Physics and Center for Relativistic Astrophysics, Georgia Institute of Technology, Atlanta, GA 30332, USA
[6]Dept. of Physics, Southern University, Baton Rouge, LA 70813, USA
[7]Dept. of Physics, University of California, Berkeley, CA 94720, USA
[8]Lawrence Berkeley National Laboratory, Berkeley, CA 94720, USA
[9]Institut für Physik, Humboldt-Universität zu Berlin, D-12489 Berlin, Germany
[10]Fakultät für Physik & Astronomie, Ruhr-Universität Bochum, D-44780 Bochum, Germany
[11]Physikalisches Institut, Universität Bonn, Nussallee 12, D-53115 Bonn, Germany
[12]Université Libre de Bruxelles, Science Faculty CP230, B-1050 Brussels, Belgium
[13]Vrije Universiteit Brussel, Dienst ELEM, B-1050 Brussels, Belgium
[14]Dept. of Physics, Chiba University, Chiba 263-8522, Japan
[15]Dept. of Physics and Astronomy, University of Canterbury, Private Bag 4800, Christchurch, New Zealand
[16]Dept. of Physics, University of Maryland, College Park, MD 20742, USA
[17]Dept. of Physics and Center for Cosmology and Astro-Particle Physics, Ohio State University, Columbus, OH 43210, USA
[18]Dept. of Astronomy, Ohio State University, Columbus, OH 43210, USA
[19]Dept. of Physics, TU Dortmund University, D-44221 Dortmund, Germany
[20]Dept. of Physics, University of Alberta, Edmonton, Alberta, Canada T6G 2E1
[21]Département de physique nucléaire et corpusculaire, Université de Genève, CH-1211 Genève, Switzerland
[22]Dept. of Physics and Astronomy, University of Gent, B-9000 Gent, Belgium
[23]Dept. of Physics and Astronomy, University of California, Irvine, CA 92697, USA
[24]Laboratory for High Energy Physics, École Polytechnique Fédérale, CH-1015 Lausanne, Switzerland
[25]Dept. of Physics and Astronomy, University of Kansas, Lawrence, KS 66045, USA
[26]Dept. of Astronomy, University of Wisconsin, Madison, WI 53706, USA
[27]Dept. of Physics and Wisconsin IceCube Particle Astrophysics Center, University of Wisconsin, Madison, WI 53706, USA
[28]Institute of Physics, University of Mainz, Staudinger Weg 7, D-55099 Mainz, Germany
[29]Université de Mons, 7000 Mons, Belgium
[30]T.U. Munich, D-85748 Garching, Germany
[31]Bartol Research Institute and Department of Physics and Astronomy, University of Delaware, Newark, DE 19716, USA
[32]Dept. of Physics, University of Oxford, 1 Keble Road, Oxford OX1 3NP, UK
[33]Dept. of Physics, University of Wisconsin, River Falls, WI 54022, USA
[34]Oskar Klein Centre and Dept. of Physics, Stockholm University, SE-10691 Stockholm, Sweden
[35]Department of Physics and Astronomy, Stony Brook University, Stony Brook, NY 11794-3800, USA
[36]Dept. of Physics and Astronomy, University of Alabama, Tuscaloosa, AL 35487, USA
[37]Dept. of Astronomy and Astrophysics, Pennsylvania State University, University Park, PA 16802, USA
[38]Dept. of Physics, Pennsylvania State University, University Park, PA 16802, USA
[39]Dept. of Physics and Astronomy, Uppsala University, Box 516, S-75120 Uppsala, Sweden
[40]Dept. of Physics, University of Wuppertal, D-42119 Wuppertal, Germany
[41]DESY, D-15735 Zeuthen, Germany
[a]Physics Department, South Dakota School of Mines and Technology, Rapid City, SD 57701, USA
[b]Los Alamos National Laboratory, Los Alamos, NM 87545, USA
[c]also Sezione INFN, Dipartimento di Fisica, I-70126, Bari, Italy
[d]Department of Physics, Sungkyunkwan University, Suwon 440-746, Korea
[e]NASA Goddard Space Flight Center, Greenbelt, MD 20771, USA

**Acknowledgements**

We acknowledge the support from the following agencies: U.S. National Science Foundation-Office of Polar Programs, U.S. National Science Foundation-Physics Division, University of Wisconsin Alumni Research Foundation, the Grid Laboratory Of Wisconsin (GLOW) grid infrastructure at the University of Wisconsin - Madison, the Open Science Grid (OSG) grid infrastructure; U.S. Department of Energy, and National Energy Research Scientific Computing Center, the Louisiana Optical Network Initiative (LONI) grid computing resources; Natural Sciences and Engineering Research Council of Canada, WestGrid and Compute/Calcul Canada; Swedish Research Council, Swedish Polar Research Secretariat, Swedish National Infrastructure for Computing (SNIC), and Knut and Alice Wallenberg Foundation, Sweden; German Ministry for Education and Research (BMBF), Deutsche Forschungsgemeinschaft (DFG), Helmholtz Alliance for Astroparticle Physics (HAP), Research Department of Plasmas with Complex Interactions (Bochum), Germany; Fund for Scientific Research (FNRS-FWO), FWO Odysseus programme, Flanders Institute to encourage scientific and technological research in industry (IWT), Belgian Federal Science Policy Office (Belspo); University of Oxford, United Kingdom; Marsden Fund, New Zealand; Australian Research Council; Japan Society for Promotion of Science (JSPS); the Swiss National Science Foundation (SNSF), Switzerland.

# An improved data acquisition system for supernova detection with IceCube

THE ICECUBE COLLABORATION[1],

[1]*See special section in these proceedings*

*david.heereman@ulb.ac.be*

**Abstract:** With an array of 5160 photomultiplier tubes, IceCube monitors one cubic kilometer of deep Antarctic ice at the geographic South Pole. Neutrinos are detected via the Cherenkov photons emitted by charged secondaries from their interactions in matter. Due to low ice temperatures, the photomultipliers dark noise rates are particularly low. Therefore a collective rate enhancement introduced by interacting neutrinos in all photomultipliers can be used to search for the signal of galactic core collapse supernovae, even though each individual neutrino interaction is sub-threshold for forming a trigger. At present, rates of individual photomultipliers are recorded in 1.6384 ms intervals which limits the time resolution and does not allow to exploit signal correlations between the sensors. An extension to the standard data acquisition, called HitSpooling, overcomes these limitations by buffering the full raw data stream from the photomultipliers for a limited time. Thus, the full set of data can be analyzed when a supernova occurs, allowing for the determination of the average neutrino energy and the analysis of the fine time structure of the neutrino light curve. The HitSpooling system will also significantly help in understanding the noise behavior of the detector and reduce the background induced by atmospheric neutrinos to the supernova analysis.

**Corresponding authors**:
David Heereman[1], Volker Baum[2], Ronald Bruijn[3]

[1] *Université Libre de Bruxelles, 1050 Brussels, Belgium*
[2] *Institute of Physics, Johannes Gutenberg University of Mainz, D-55099 Mainz, Germany*
[3] *Laboratory for High Energy Physics , 1015 Lausanne, Switzerland*

## 1 Introduction

IceCube is a neutrino telescope based in the Antarctic ice at the geographic South Pole and consists of 86 strings, each equipped with 60 photomultiplier tubes (PMTs). It detects neutrinos via the Cherenkov photons emitted by charged secondaries from their interactions in matter. This unique detector monitors a cubic kilometer of cold and inert ice at depths between 1450 m and 2450 m. Each of its 5160 digital optical modules (DOMs) houses a 17 inch photomultiplier tube, as well as several electronics boards containing a processor, memory, flash file system and realtime operating system that allows each DOM to operate as a complete and autonomous data acquisition system. For most of IceCube, the DOMs are deployed in the ice with 17 m (125 m) vertical spacing between DOMs (horizontal spacing between strings), while in the DeepCore volume the DOMs are deployed in a denser spacing of 7 m distance on a string and 72 m between strings. The DeepCore DOMs are also equipped with high quantum efficiency PMTs that are roughly 35% more efficient than standard PMTs.
For supernova neutrinos with energy $\mathcal{O}(10\,\mathrm{MeV})$, the dominant interaction in ice or water is the inverse beta process ( $\overline{\nu}_e + p \rightarrow e^+ + n$ ), leading to positron tracks of about

$$0.56\,\mathrm{cm} \cdot \frac{E_\nu}{[\mathrm{MeV}]} \qquad (1)$$

Monte Carlo studies yield an average number of 178 photons per MeV energy of the positron, considering only a range of wavelengths from 300 nm to 600 nm accessible to our optical modules. For $10^7$ expected neutrinos from a supernova at 10 kpc distance we assume to detect up to 3.6 million neutrino induced PMT hits. As elaborated in [1, 2], neutrino detectors in general and IceCube especially due to its unique size are well suited to monitor our galaxy for supernovae. This detection principle is realized in IceCube by monitoring the count rates of individual DOMs in ms time bins. This *scaler* data stream is decoupled from the *hit* data stream that holds more detailed information, but is only saved for events that trigger and may not be available in the case of a supernova.
Motivated by the goal to gain as much information as possible in case of a galactic core-collapse supernovae, a new data buffer, called *HitSpooling* and first mentioned in [3], is now realized in IceCube. In combination with a newly implemented interface, the separation of the *scalar* data used for supernova searches and the *hit* data stream used for other physics searches can now be overcome. We introduce HitSpooling and the implementation of an interface between the two existing data streams. Furthermore we discuss the physics capabilities that will become accessible through the HitSpooling system.

## 2 Data Streams in IceCube

IceCube's data streams are handled by two data acquisition systems (DAQs). The supernova system is responsible for processing and analyzing hit counts (*scaler*) data and the standard DAQ that takes care of waveforms (*hits*).

## 2.1 Hits Data Stream

Whenever the anode of the PMT in a DOM exceeds a discriminator threshold of 0.25 photo-electrons[1], a *hit* is recorded.

In order to reduce data transfer rates, hardware trigger signals are exchanged between neighboring DOMs to look for certain coincidence conditions, so called soft or hard local coincidences (*SLC* or *HLC*), respectively. In HLC mode, the triggered DOM looks for hits in next-neighbor and next-to-next-neighbors within a time-window of $\pm 1$ µs. If this condition is met, an HLC is present and all recorded data from both DOMs will be transferred to the surface. In SLC, also isolated DOMs with no neighboring hit DOM will be processed, a time-stamp will be assigned to it, but only information from the bins around the waveform peak will be transferred to the surface. In summary, the hit record holds a timestamp, a 3-bin peak- or 128-bin full-measure of the waveform, depending on trigger conditions (SLC or HLC), as well as trigger information [4]. There are several DAQ elements involved from capturing, aggregating to transferring the data from a hit:

1. The *DOM* digitizes the waveform on-board in a time-interval of 6.4 µs which generously covers the maximal time that light from the most energetic events is expected to arrive in any DOM.
2. The *DOMHub*, located in the surface laboratory, is an industrial computer that communicates with all DOMs on one string.
3. The *StringHub* is a software element that runs on the DOMHub and collects hits per string as they arrive chronologically. It also converts DOM hits in physics-ready hits that are suitable for other higher level software DAQ elements like triggers and event-builders.

The hit streams from each of the 60 DOMs are fed into the StringHub, which creates trigger payloads and buffers the hits in memory for possible later readout by the event-builder. Further processing and filtering of the events is performed on a cluster of machines at the South Pole before the data is sent via satellite to the data warehouses in the Northern hemisphere.

## 2.2 Scaler Data Stream

Independent of the above mentioned coincidence conditions in the DOMs, a firmware integrated scaler adds asynchronously all discriminated PMT pulses in intervals and assigns them a timestamp. The interval length arises from adding up scalers from $2^{16}$ clock cycles per bin: $2^{16}/40\,\mathrm{MHz} = 1.6384\,\mathrm{ms}$. Since these intervals are not synchronized between the individual DOMs, the supernova DAQ collects the scaler data and re-bins them to the desired synchronized rates in global 2 ms bins [5].

Since the supernova analysis algorithm looks for subtle changes in the background rate, understanding the noise properties of the individual modules as well as the entire detector is essential. IceCube DOMs feature a very low noise rate of on average $\approx 540\,\mathrm{Hz}$ for standard DOMs, see figure 1. Due to stronger absorption and scattering of light in the dustier ice regions, a drop in the rate is visible at

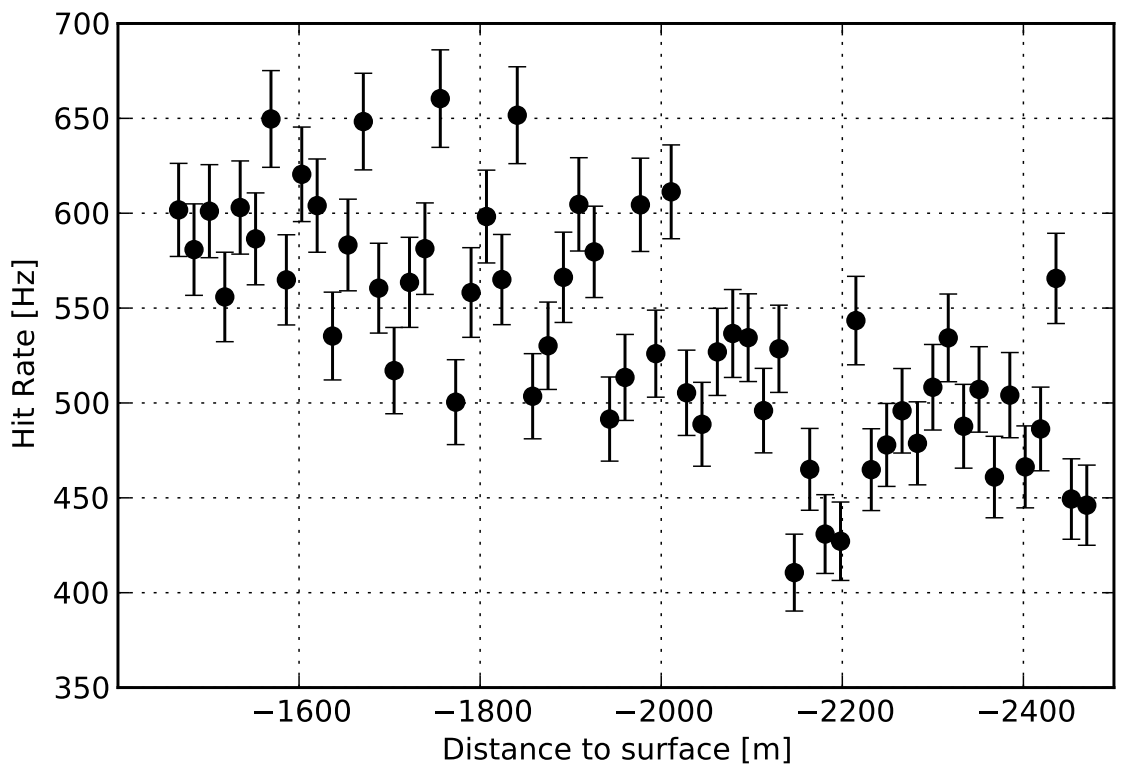

**Figure 1**: Average noise rate for each DOM on an arbitrary string in IceCube, calculated from HitSpool data. Besides PMT-to-PMT variations, depth-depending ice characteristics, such as temperature and dust-concentration, are the main contribution to the scattering in the data points.

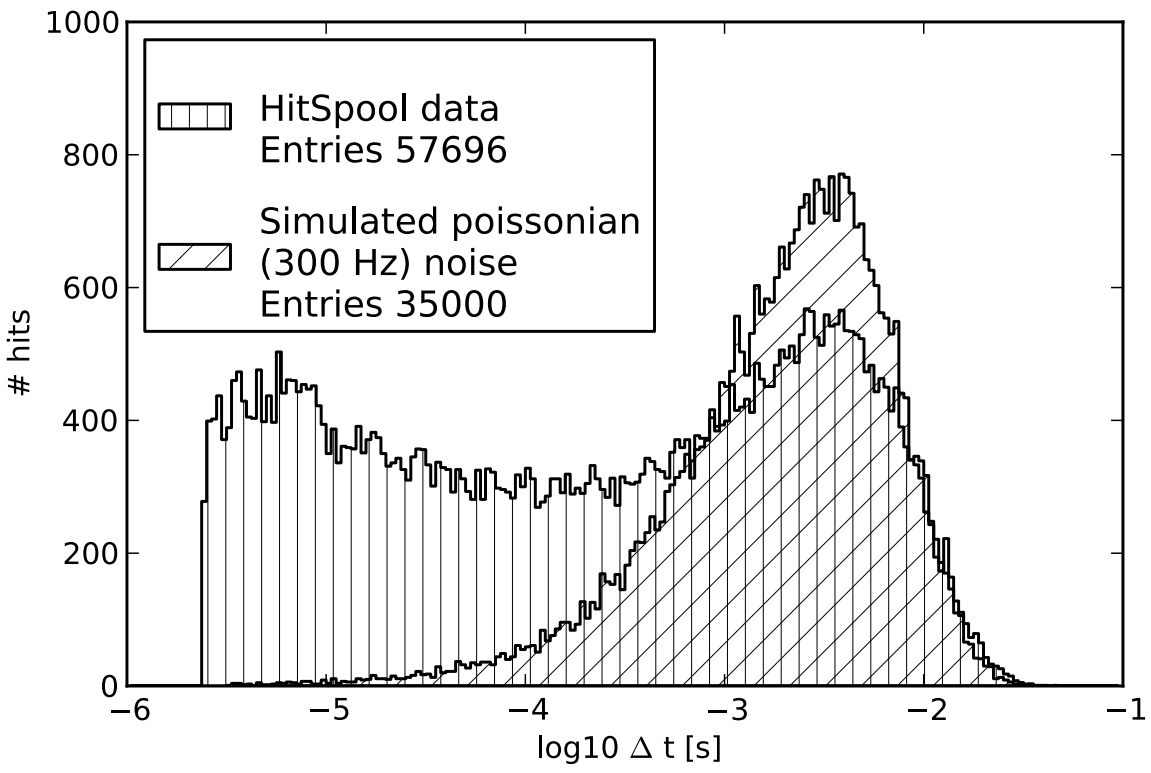

**Figure 2**: Logarithmic time distribution of successive hits in a single DOM. Contributions by uncorrelated noise hits from radioactive decays are simulated with 300 Hz.

around 2000 m depth. The hit rate is partly originating from uncorrelated Poissonian noise contributions by radioactivity, thermal noise and atmospheric muons, where the latter are considered as background in this analysis. A second contribution is due to correlated noise from Cherenkov radiation and/or scintillation in the glass of the PMT and the pressure sphere of the DOM, see figure 2 . Correlated noise hits can last for $\mathscr{O}(100\,\mu\mathrm{s})$ and can be suppressed by applying an artificial deadtime $t_{dead}$ of 250 µs to every hit in the DOM. In this way, the largest part of the non-Poissonian behavior of the background noise is eliminated and the remaining noise rate of $285 \pm 26\,\mathrm{Hz}$ is very stable in time and only slightly varying with depth [6]. This reduced scaler stream is the raw data input of the supernova DAQ, see figure 3. As a first step, the supernova DAQ calculates individual noise rates for each DOM via counting scalers for a given time-interval. For sufficiently large time-intervals, these individual noise rates follow a log-normal

1. Photo-electron voltage is defined as the anode voltage level in the PMT that's caused by a single photo-electron after being amplified through all diode stages in the photomultiplier.

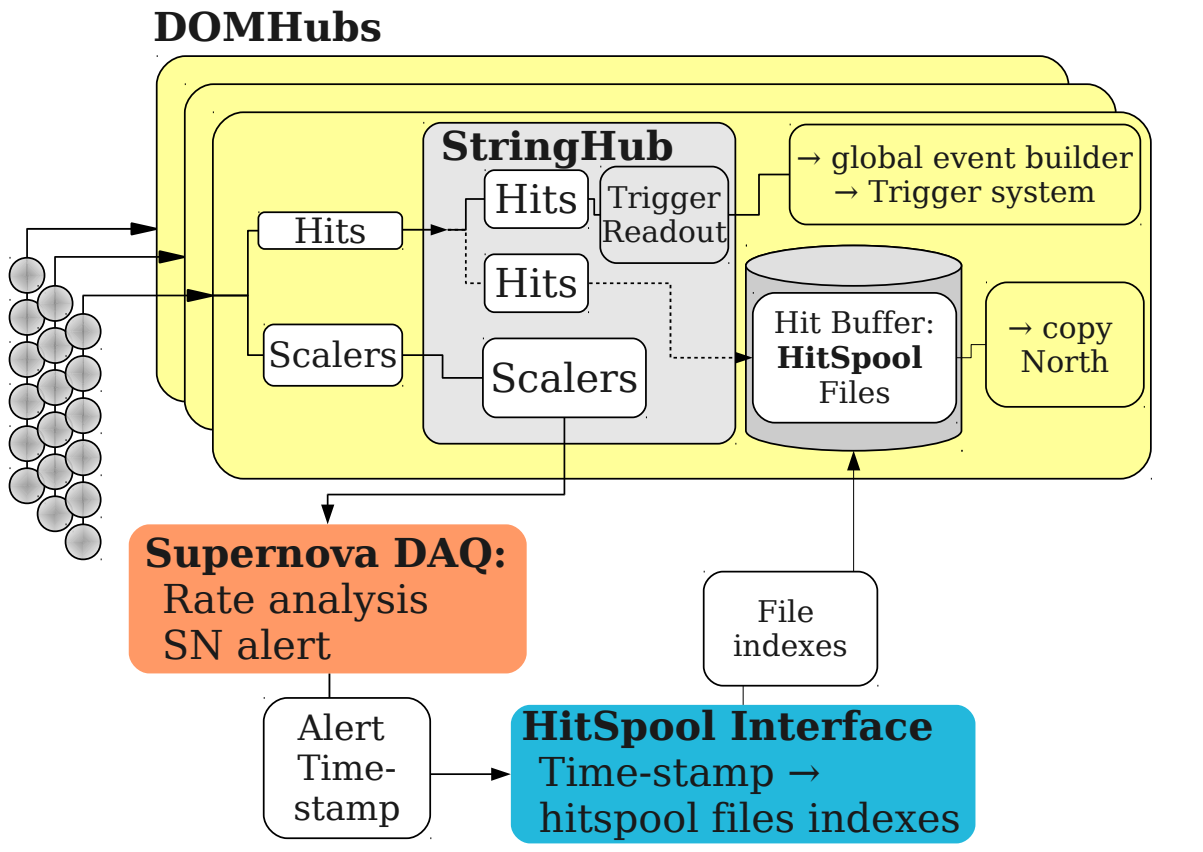

**Figure 3**: Block-diagram of components of data streams in IceCube. The new HitSpool raw data stream is shown by the dashed line.

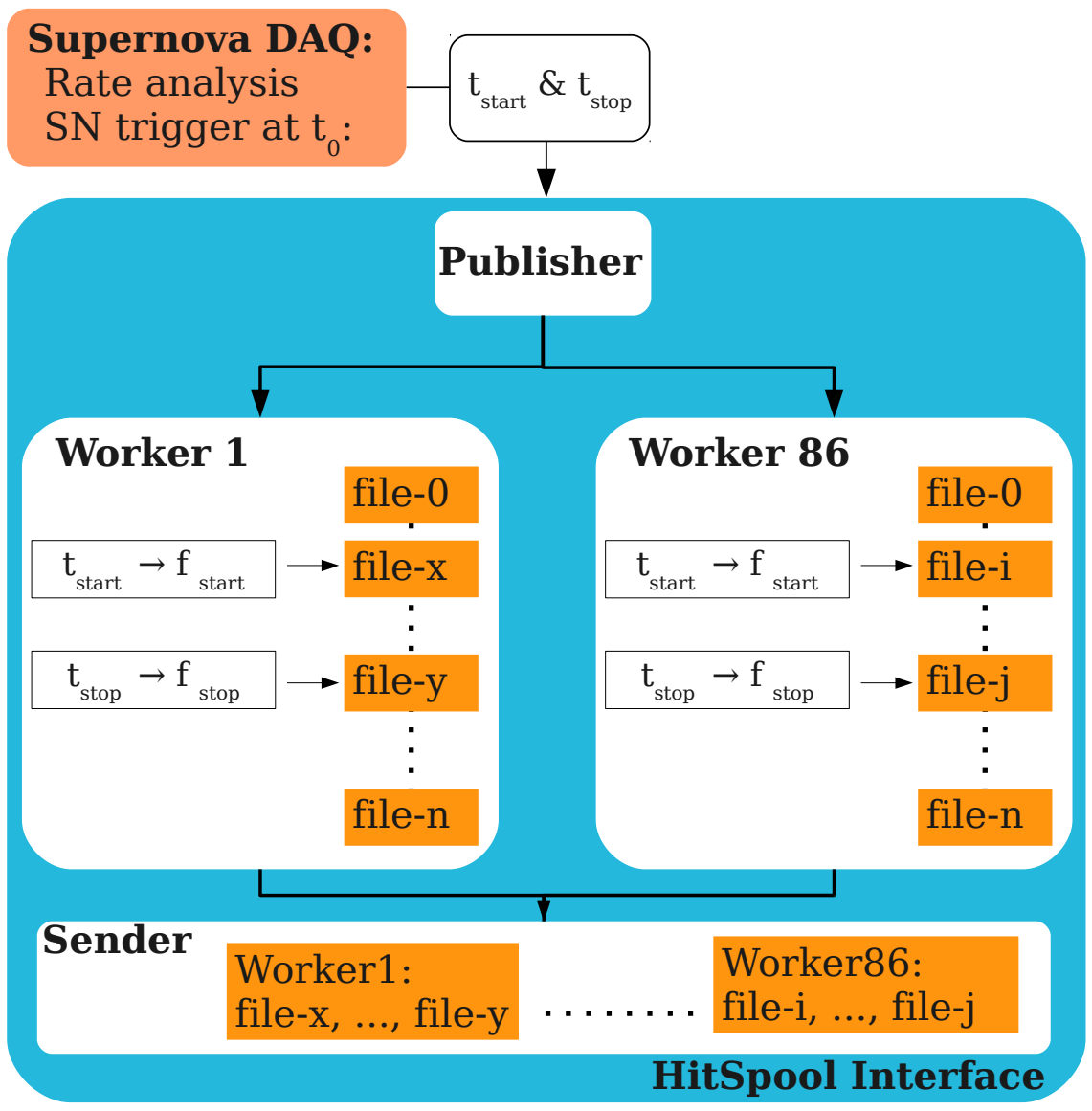

**Figure 4**: Block-diagram of components of HitSpool Interface. HitSpool files from all 86 Strings are transferred in parallel.

distribution which can be approximated for convenience by a Gaussian distribution that returns expectation values and corresponding standard deviation expectation values. The expectation values are calculated from moving 300 s time-intervals that leave out a $[-30\,\mathrm{s}, +30\,\mathrm{s}]$ window around the investigated time $t_0$, in order to reduce the impact of a long-lasting signal on the mean rates. Based on a maximum likelihood analysis, the most likely *collective rate deviation*, $\Delta\mu$, of all DOM noise rates can be evaluated (see [5] for details). The supernova DAQ issues an alert when $\xi > 7.3$ where $\xi$ is the significance and calculated as $\xi = \Delta\mu/\sigma_{\Delta\mu}$, with $\sigma_{\Delta\mu}$ the uncertainty of the collective rate deviation $\Delta\mu$. In case a trigger occurs, which happens roughly every two weeks, an alarm including real-time datagrams is sent to the Supernova early warning system (SNEWS) [7]. Furthermore, a message is sent to the HitSpool interface (see section 3) requesting $[-30\,\mathrm{s}, +60\,\mathrm{s}]$ around the trigger time-stamp $t_0$ from the hitspool data stream, which is explained in the next section.

### 2.3 HitSpool Stream

In addition to the previously described hits data stream that is stored in memory, a copy of that raw data is buffered to disk in the DOMHub, see figure 3. HitSpooling accumulates on average 2 MB of raw data per string per second which is stored in files of variable duration, default is 15 seconds. The hits are buffered in a spooling cycle which overwrites itself after a given time, currently up to 16 hours, depending on the disk size. As the HitSpooling is independent of any other StringHub element it also could increase the uptime of the detector. HitSpooling also serves as a worst case scenario backup for nearby supernovae when IceCube's DAQ systems could saturate due to the massive amount of hits. In order to keep track of the spooling process on a string, a text file is written inside the buffer directory that holds the following information:

- Time stamp of first hit on string at run start: $t_{run}$
- Interval, e.g. length of each HitSpool file: $t_{ival}$
- Time stamp of first hit in current file: $t_{cur}$
- File index of currently active HitSpool file: $n_{cur}$
- Max number of files per HitSpool cycle: $n_{max}$

From these five values it is possible to calculate all necessary information about the hits contained in a HitSpool file without accessing the file itself and in this way avoiding any interference with the actual data taking.

## 3 HitSpool Interface

The HitSpool interface, see figure 4, is the centerpiece for accessing hitspool data. It consist of several services running on the South Pole on several machines, communicating via the *ØMQ* messaging service [8]. Whenever the supernova DAQ triggers a supernova candidate at $t_0$, a message is sent to the HitSpool interface including two time-stamps, $t_{start} = t_0 - 30\,s$ and $t_{stop} = t_0 + 60\,s$. The *Publisher* service is running on the central machine of IceCube and pushes the request from the supernova DAQ to all hubs. At the hubs, the *Worker* service calculates the HitSpool files integer indexes $f_{start}$ and $f_{stop}$ that contain $t_{start}$ and $t_{stop}$, respectively:

$$f_{start} = \frac{t_{start} - t_{run}}{t_{ival}} \mathrm{mod}(n_{max}) \qquad (2)$$

$$f_{stop} = \frac{t_{stop} - t_{run}}{t_{ival}} \mathrm{mod}(n_{max}), \qquad (3)$$

where $\mathrm{mod}(n_{max})$ accounts for the fact that HitSpooling overwrites existing files when the first loop is finished after time $t = n_{max} \cdot t_{ival}$. The data is transferred to a central location where all hubs' data is collected. The post-processing *Sender* service is responsible for sending data to the North.

## 4 Physics Capabilities

HitSpooling is not only of value for noise studies, as shown in figure 1, but also contributes to the supernova search with IceCube. HitSpooling provides the possibility to study any arbitrary time-interval with which the start time of the supernova can be determined. Furthermore, recent astrophysical models [9] predict high frequency modulations in the neutrino flux, e.g. from the collapse of fast rotating stars. In case of a galactic supernova, the event rates would be high enough that these can be observed in the HitSpool data. The main profit from the HitSpool data stream arises from a better muon subtraction and the possibility to estimate the average neutrino energy. Both aspects will be explained in more detail in the next sections.

### 4.1 Muon Subtraction

As discussed in [3], roughly half of the hits in the detector introduced by atmospheric muons can be subtracted off-line already by identifying triggered muons in the standard data stream. Atmospheric muons are detected in Icecube at an average rate of 16 Hz. For the supernova analysis, these are a cause of correlated noise which artificially widens the significance distribution of the noise rate fluctuations by a factor of $\approx 1.45$ relative to the expectations from pure poissonian noise.
By removing hits associated with muon triggers in IceCube, the broadening of the significance distribution is reduced [10]. Muon subtraction will be improved by HitSpooling data, where one is able to associate also isolated hits to the atmospheric muon introduced hits that where not counted in the muon triggered events. This could be done by defining a radius $R$ and a time-window $t$ around muons induced HLC hits and subtract these from the HitSpool data set. Furthermore, standard tools like track-reconstructions can be used to identify muon signal related hits. Both techniques are under investigation and are expected to improve the reduction of atmospheric muons. Such cleaned data are expected to narrow the significance distribution and remove non-Gaussian tails which will increase the supernova detection range of IceCube.

### 4.2 Energy Estimation

As shown in equation 1, the track length and thus the probability to detect a photon from a supernova neutrino interaction in the ice increases with the energy. Therefore the probability to detect light from a single interaction in multiple DOMs also rises with the neutrino energy. A method developed by [11], called the *coincident hit method*, takes advantage of the HitSpool data stream by investigating coincident hits in one, two and three DOMs in a time-window of 150 ns and a spatial distance up to next-to-next neighboring DOMs on next-to-nearby strings. The various hit patterns are sensitive to different regions of the energy spectrum depending on the patterns geometry [3].
Evaluating the ratio of, e.g. the two-fold nearest-neighbor DOM hit rate over the single DOM hit rate, one obtains an energy dependent observable. Assuming a specific shape of the neutrino spectrum, this observable can be used to determine the absolute energy scale of the spectrum (i.e. determine the mean neutrino energy) [12].
It is expected that this method puts a lower bound of 30% on the energy resolution for a core-collapse supernova at 10 kpc distance assuming a Hüdepohl model, see figure 5. For other models with heavier stars and harder neutrino spectra this resolution will improve. It is planned for the near future to test this algorithm on HitSpool data.

## 5 Summary & Conclusion

The HitSpool data stream is a powerful tool that helps improving the supernova detection within IceCube by investigating the noise properties of the detector and by identifying atmospheric muons that did not trigger the detector. An analysis of the coincident hit method will give a handle on the estimation of the average neutrino energy. In order to access HitSpooling data, an interface was implemented and is running stably at the South Pole to retrieve 90 seconds of raw detector data, and sends these automatically via satellite to the North, in case of a high significance supernova trigger candidate. Furthermore, HitSpooling is of general benefit for externally triggered events, e.g. Gamma Ray Bursts, not only for the supernova system.

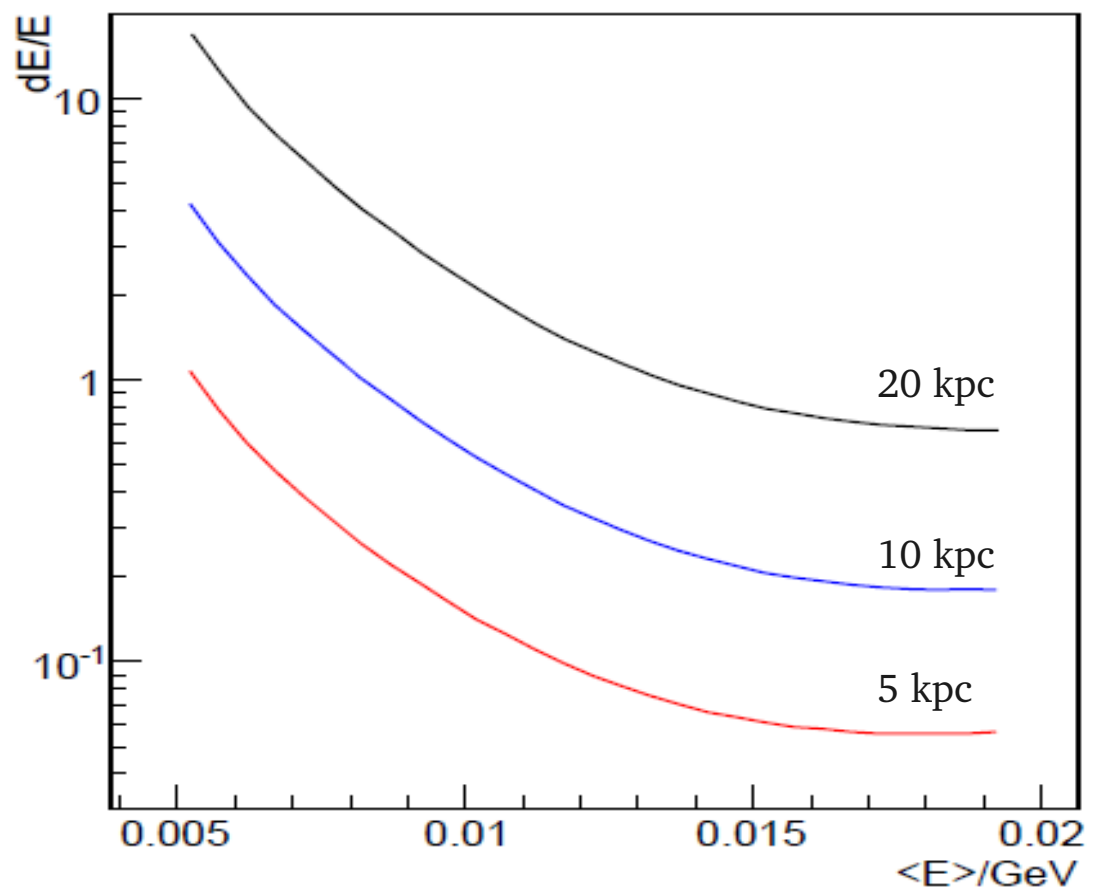

**Figure 5**: Energy resolution, assuming a time integrated 8.8 $M_{\odot}$ O-Mg-Ne model with a shape parameter $\alpha = 2.84$ for a supernova happening at 5 kpc, 10 kpc and 20 kpc [12]

# On the detection of galactic core collapse supernovae with IceCube

THE ICECUBE COLLABORATION[1],

[1] *See special section in these proceedings*

*volker.baum@icecube.wisc.edu*

**Abstract:** The IceCube Neutrino Observatory, situated at the geographic South Pole, was mainly designed to detect energies greater than 100 GeV with its lattice of 5160 photomultiplier tubes monitoring 1 km$^3$ of clear Antarctic ice. Neutrinos undergoing interactions produce charged secondaries that in turn produce Cherenkov photons, whose arrival time are recorded in the photomultipliers. Due to subfreezing ice temperatures, the photomultiplier dark noise rates are particularly low. This allows IceCube to extend its searches to several second long bursts of $\mathcal{O}$(10 MeV) neutrinos expected to be emitted from galactic core collapse supernovae. By observing a collective rise in all photomultiplier rates, IceCube would provide the highest statistical precision for close-by supernova. In this paper, the method to determine a conservative upper limit on the number of core collapse supernovae in our galaxy is presented, based on IceCube data taken from April 5, 2008 to May 13, 2011, in configurations with 40, 59 and 79 strings deployed. Systematic uncertainties, mainly due to unknown ice properties, are assessed with a GEANT based Monte Carlo.

**Corresponding authors:** Gösta Kroll[2], Benedikt Riedel[3], Volker Baum[2]
[2] *Institute of Physics, University of Mainz, Germany*
[3] *Department of Physics, University of Wisconsin, WI, USA*

## 1 Introduction

The rate of galactic stellar collapses, including those obscured in the optical, is estimated [1] to be in the range of $(1.7-2.5)$ per 100 years. The best experimental upper limit from the absence of a neutrino signal, calculated in a specific SN model, is $< 9$ per 100 years at the 90% C.L. [2].

IceCube, a grid of 5160 photo sensors embedded in the ice of the antarctic glacier, is uniquely suited [3] to monitor our Galaxy for supernovae due to its 1 km$^3$ size and its location. In the inert and $-43\,^\circ$C to $-20\,^\circ$C cold ice, IceCube's photomultiplier noise rates average around 540 Hz. At depths between (1450 – 2450) m, the detector is partly shielded from cosmic ray muons. The inverse beta process $\bar{\nu}_\mathrm{e} + \mathrm{p} \rightarrow \mathrm{e}^+ + \mathrm{n}$ dominates supernova neutrino interactions with $\mathcal{O}(10\,\mathrm{MeV})$ energy in ice, with total lepton tracks lengths of about $0.56\,\mathrm{cm} \cdot E_\nu/\mathrm{MeV}$ along which $178 \cdot E_{\mathrm{e}^+}/\mathrm{MeV}$ Cherenkov photons are radiated in the (300 – 600) nm wavelength range. Due to their low cross section in water, electron-neutrinos contribute less than 5% to the rate. From the approximate $E_\nu^2$ dependence of the cross section and the linear energy dependence of the track length, the light yield per neutrino roughly scales with $E_\nu^3$. The detection principle was demonstrated with the AMANDA experiment, IceCube's predecessor [4]. With absorption lengths exceeding 100 m, photons travel long distances in the ice so that each DOM effectively monitors several hundred cubic-meters of ice. Typically, only a single photon from each interaction reaches one of photomultipliers that are vertically (horizontally) separated by roughly 17 m (125 m). The DeepCore subdetector, equipped with a denser array of high efficiency photomultipliers, provides higher detection and coincidence probabilities.

Although the rate increase in individual light sensors is not statistically significant, the effect will be clearly seen once the rise is considered collectively over many sensors. IceCube is the most precise detector for analyzing the neutrino light curve of close supernovae [5]. Since 2009, IceCube has been sending real-time datagrams to the Supernova Early Warning System (SNEWS) [6] when detecting supernova candidate events.

The supernovae search algorithms employed in the paper are based on count rates of individual optical modules stored in 1.634 ms time bins. By buffering the full photomultiplier raw data stream that is stored around supernova candidate triggers, additional information, e.g. on the average neutrino energy, can be retrieved [7, 8].

## 2 Effective volume for supernova detection

The signal hit rate per DOM for the inverse beta decay is given by

$$\begin{aligned} R(t) &= \epsilon_\mathrm{deadtime} \frac{n_\mathrm{target} L^\nu_\mathrm{SN}(t)}{4\pi d^2 \overline{E_\nu}(t)} \int_0^\infty dE_{\mathrm{e}^+} \int_0^\infty dE_\nu \\ &\times \frac{d\sigma}{dE_{\mathrm{e}^+}}(E_{\mathrm{e}^+}, E_\nu)\, V^\mathrm{eff}_{\mathrm{e}^+} f(E_\nu, \overline{E_\nu}, \alpha_\nu, t)\,, \end{aligned} \quad (1)$$

where $n_\mathrm{target}$ is the density of targets in ice, $d$ is the distance of the supernova, $L^\nu_\mathrm{SN}(t)$ its luminosity, and $f(E_\nu, \overline{E_\nu}, \alpha_\nu, t)$ is the normalized $E_\nu$ distribution depending on a shape parameter $\alpha_\nu$ and the average neutrino energy $\overline{E_\nu}$. $E_{\mathrm{e}^+}$ denotes the energy of positrons emerging from the neutrino reaction. The effective volume for a single positron, $V^\mathrm{eff}_{\mathrm{e}^+} \propto E_{\mathrm{e}^+}$, strongly varies with the photon absorption but shows little dependence on photon scattering. An artificial deadtime of $\tau = 250\ \mu$s was introduced to suppress time correlated supra-Poissonian photomultiplier pulses at low temperatures, leading to an inefficiency parametrized by $\varepsilon_{\text{deadtime}} \approx 0.87/(1 + r_{SN} \cdot \tau)$, where $r_{SN}$ denotes the rate per optical module that arises from supernova neutrinos.

A GEANT-4 based simulation of the interaction of individual supernova neutrinos in the ice and a computationally optimized tracking [9] of individual Cherenkov photons that can be run on graphical processing units, was used to determine IceCube's effective volume for supernova detection. Calibration measurements with light sources in the ice and a dust logger [10] with $< 1$ cm vertical resolution allow one to fit the depth, position and angular dependent photon absorption and scattering lengths of the ice [11, 12]. The uncertainties in these measurement lead to a range of ice models. Other important uncertainties arise from the photon tracking in the presence of Mie scattering, optical module sensitivities, as well as from cross section uncertainties which are sizeable for interactions with $^{16}$O and $^{18}$O. The effective volume per optical module was determined by injecting $1.4 \times 10^9$ positrons of 10 MeV energy with random directions and random positions inside a sphere with radius 250 m around every optical module along a string. Fig. 1 shows $V^{\text{eff}}_{\text{e}^+}/E_{\text{e}^+}$, determined from the fraction of positrons that generated photoelectrons at the cathode surface, as function of depth. The $\approx 35\%$ higher quantum efficiency of the optical modules in the high density DeepCore subdetector, installed in two ice regions below and above the main dust layer, is apparent. The effective volume scales linearly with the optical module sensitivities. Systematical uncertainties due to variations in the optical absorption and scattering lengths and the photon propagation algorithm can be deduced from the lower panels. While different descriptions of the optical ice properties show clear differences in the depth dependence of the absorption, the averaged effective volumes differ only by 7.4%. In addition, a 10% uncertainty on the average photosensor sensitivity has to be taken into account. Table 1 lists the most important detection related systematic uncertainties.

**Table 1**: Major detection related systematic uncertainties.

| Source | estimated uncertainty [%] |
|---|---|
| effective volume uncertainty | 12 |
| cross section uncertainties | 3 |
| effect of artificial deadtime | 3 |
| positron track length | 5 |

## 3 Search for galactic supernovae

The neutrino emission from a core collapse supernova strongly depends on the progenitor mass and type, leading to an order of magnitude variation of detected rates. Different methods to estimate the minimum initial mass that can produce a supernova have converged [13] to $(8 \pm 1)\ M_\odot$, where $M_\odot$ is the mass of the Sun. For these low masses, the collapse is induced by electron capture in a degenerate O-Ne-Mg core. The observed rates also depend on the assumed neutrino mass hierarchy, the influence of matter induced oscillations in the Earth ($< 8\%$), and the assumed progenitor distribution in Milky Way. It is therefore important to specify which models have been assumed for a supernova search with a neutrino detector. Three spherically symmetric models, encompassing the range of models and covering the full range of neutrino emission, were selected as benchmarks in this analysis:

The collapse of a 8.8 $M_\odot$ O-Ne-Mg progenitor to the completed formation of the deleptonized neutron star is the only examples so far, where one-dimensional simulations [14] obtain neutrino-powered supernova explosions ("Hüdepohl model"). This low mass model, with total emitted energy of $1.7 \times 10^{53}$ erg and $E(\bar{\nu}_\text{e}) \approx 12.9$ MeV, also represents a conservative lower limit.

The Lawrence-Livermore simulation [15], modeled after SN 1987A, assumes a 20 $M_\odot$ progenitor. The total emitted energy is $2.9 \times 10^{53}$ erg, of which 16 % is carried by $\bar{\nu}_\text{e}$ with 15.3 MeV energy on average.

On the high mass side, the gravitational collapse of less than solar metallicity stars exceeding 25 $M_\odot$ will lead to a limited stellar explosion, while stars exceeding 40 $M_\odot$ are not expected to explode at all ("Black Hole model"). In both cases a black hole will develop $\mathcal{O}(1\ \text{s})$ after bounce. At this point, the neutrino emission quickly comes to an end, providing a unique signature for black hole formation [16]. For the analysis presented in this paper we assume a 30 $M_\odot$ progenitor and a hard equation of state.

More than 80% of supernovae may be obscured by dust and would thus not be optically visible [17]. The search method should therefore not depend on external information. The test statistics used to search for galactic supernovae with IceCube is the significance $\xi = \Delta\mu/\sigma_{\Delta\mu}$, where $\Delta\mu$ is the most likely collective rate deviation of all optical module noise rates from their running average. $\sigma_{\Delta\mu}$ is the corresponding uncertainty calculated from the data, thus accounting for non-Poissonian behaviour in the dark rates. The significance should be centered at zero with unit width if no correlations are present. The calculation was done in consecutive, non-overlapping 500 ms wide time intervals.

Starting with a data set corresponding to in total 1101 days, several requirements are introduced to select high quality data. Short runs (0.43%), runs taken with calibration light sources (1.68%) and runs with an imperfect detector (3.93%) are discarded. Finally, more than 99% of all 500 ms time slices in each run must have associated information on the number of atmospheric muon hits. After applying these cuts, a dataset corresponding to a clean livetime of 887 days remains on which the analysis is performed.

A fast simulation, checked to produce the same results as the GEANT Monte Carlo, was used to produce toy supernovae according to the three models discussed above. The respective rates were then added to recorded optical module noise data of the detector, appropriately sampling the three periods of data taking.

Fig. 2 shows the simulated significance distribution for the detection of a supernovae as function of distance for the Hüdepohl [14], Lawrence-Livermore [15] and Black Hole [16] models.

Fig 3 shows the distribution of significances for the 8.8 $M_\odot$ [14] and 20 $M_\odot$ [15] models, assuming two galactic progenitor distribution models [18, 19] (The significances of the Black Hole model lie beyond the range of the plot). Also shown is the measured significance distribution for 3 years of IceCube data taking.

The distribution narrows substantially, when hits from atmospheric muons are subtracted (see shaded distribution in Fig. 3). Single cosmic ray showers can produce muon bundles that trigger many optical modules. While cosmic ray muons contribute to the count rates of individual optical modules by only 3%, these hits are correlated across the detector, broadening the significance distribution as function of detector size. It also gives rise to a seasonal dependence of the trigger rate. This explains why the significance distribution in IceCube is widened by a factor of 1.26 - 1.46 compared to the expectation of unity, depending on the number of optical modules participating in the detector configuration and the season. It is possible to subtract a large fraction of the hits introduced by atmospheric muons from the total noise rate offline, as the number of hits is recorded for all triggered events. The width of the significance distribution then decreases to about 1.06, close to the expectation of unity. We use the subtracted distribution, which shows a substantially better separation between noise data and expected signal (shaded area in Fig. 3), to search for signs of core collapse supernovae in the IceCube data.

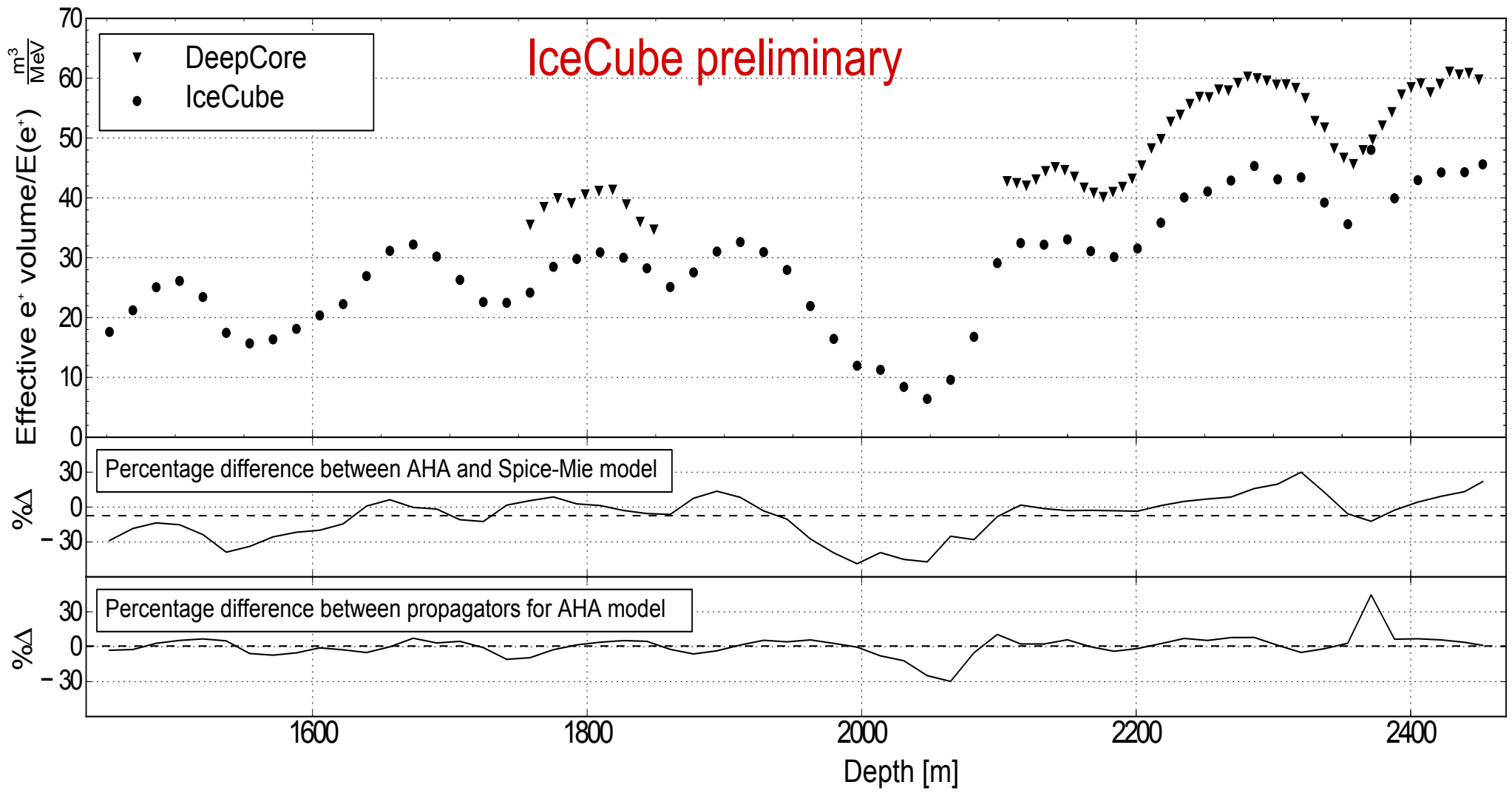

**Figure 1**: $V^{\mathrm{eff}}_{\mathrm{e}^+}/E_{\mathrm{e}^+}$ as function of depth for a recent model (Spice-Mie) of the optical absorption and scattering lengths in the antarctic glacier. The results are given for the IceCube (circles) and DeepCore (triangles) detectors. The uncertainties due to a different description of the optical ice properties (AHA model) and the way the photon propagation is implemented (tabulated vs. individual propagation of each photon) are indicated in the lower panels.

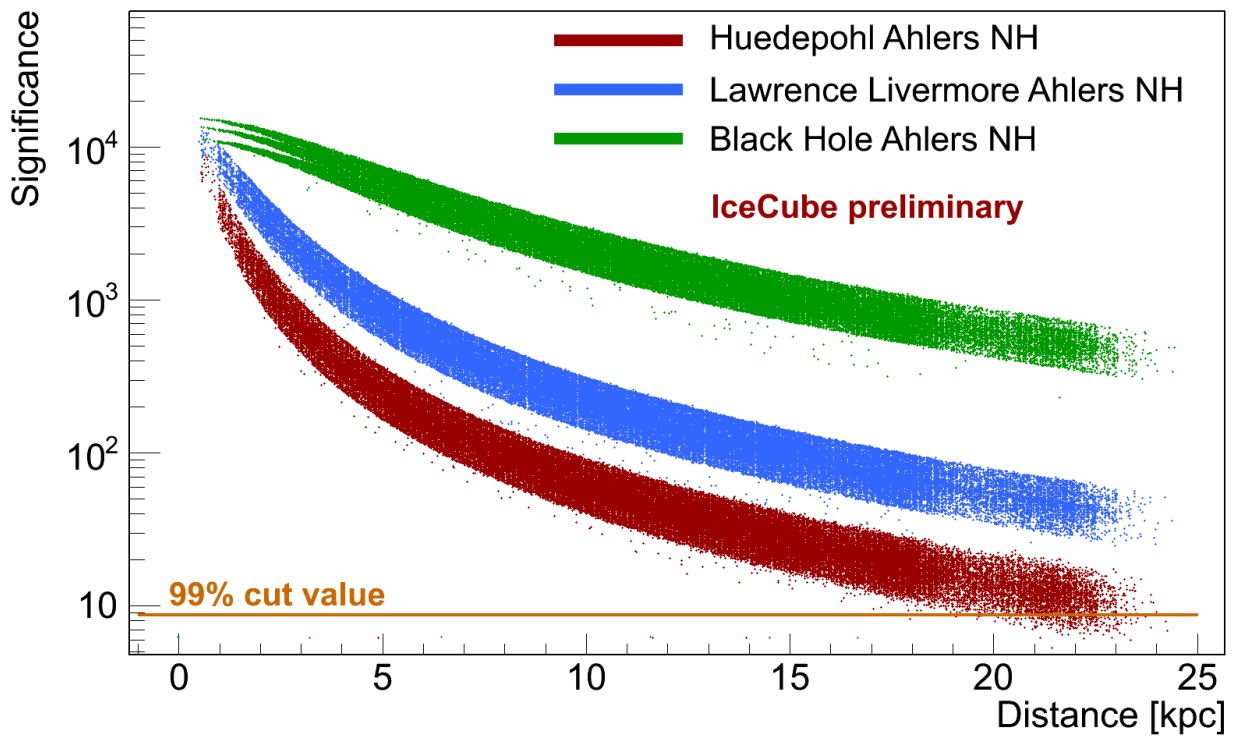

**Figure 2**: Significance versus distance for the three models described in the text. Three periods, corresponding to detector configurations with 40, 59 and 79 strings deployed, were simulated. The significance rises with the square root of the corresponding number of optical modules (particularly well visible for the Black Hole distribution at short distances). The changing density of points reflects the spiral nature of our Galaxy.

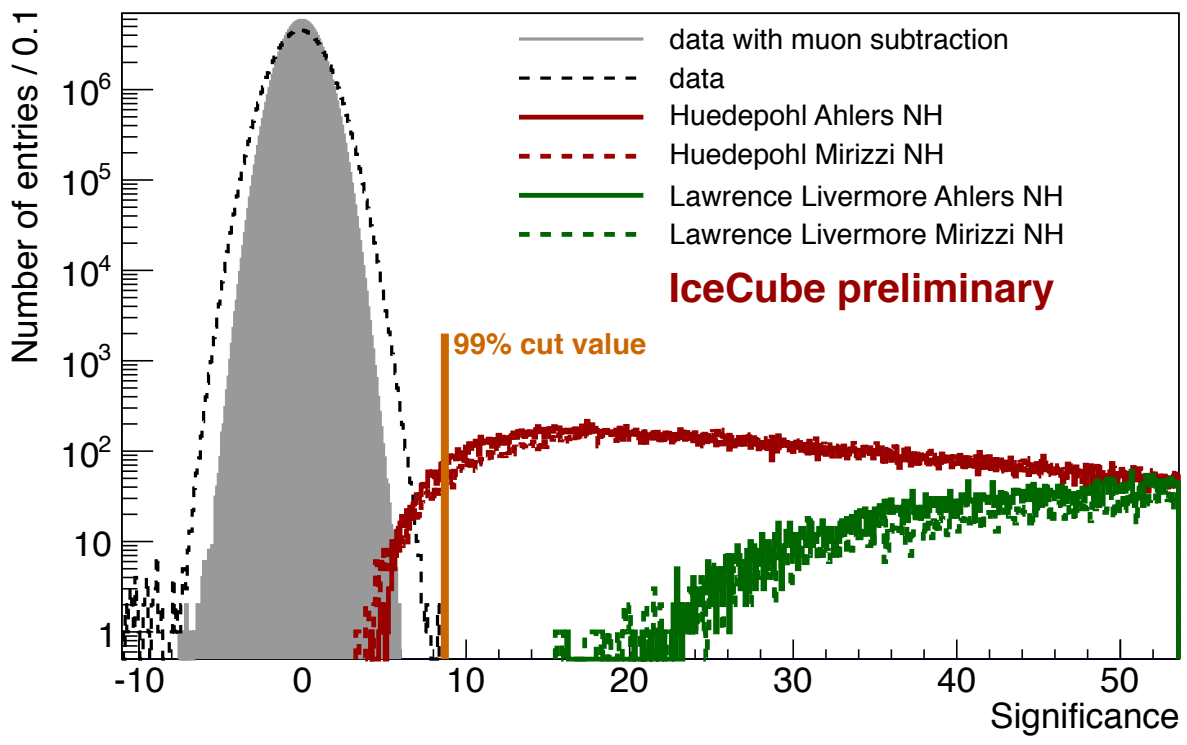

**Figure 3**: Measured significance distributions without (black line) and with (shaded grey) subtraction of atmospheric muons for data taken from April 2008 to May 2011 (significances above 6 standard deviations are kept blind). The distributions hardly overlap with the significances expected for the three supernova models studied assuming two progenitor radial distribution models for our Milky Way ([18] solid line, [19] dashed line, arbitrary normalization). Satellite and dwarf galaxies, such as the Magellanic Clouds, are not included in the simulation.

The analysis was kept blind for atmospheric muon corrected significance values above $6\,\sigma$. To be on the conservative side, we assumed a normal neutrino mass hierarchy, the progenitor distribution of [19] and lowered the effective volume by 15% to account for detection related systematic uncertainties and matter effects in the material of the Earth. We defined the significance cuts by requiring that at least 99% of all simulated core collapse supernovae in our Galaxy would be detected for the Hüdepohl model. Fig. 4 shows the fraction of supernovae missed as function of significance cut for the three models studied. Note that the choice of neutrino mass hierarchy matters only for the Lawrence-Livermore model (dotted curves).

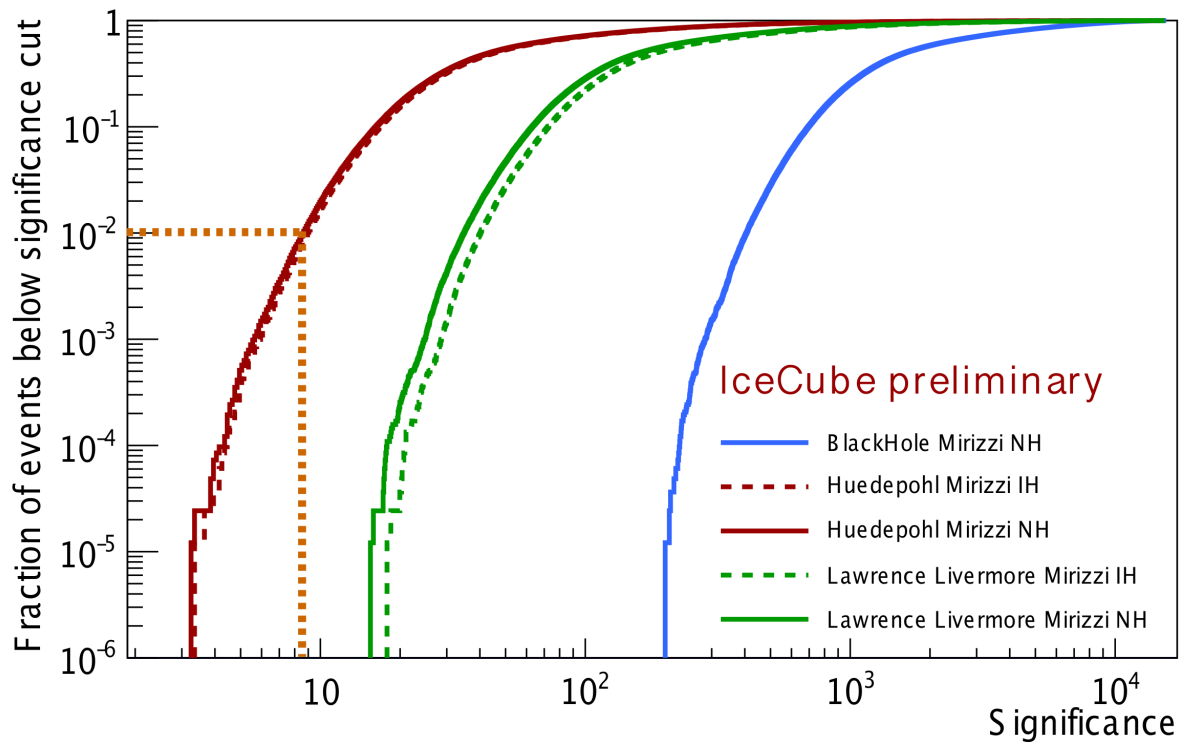

**Figure 4**: Fraction of supernovae missed in our Galaxy by imposing a cut on significance. A 15% systematic detector uncertainty is taken into account by respectively decreasing the nominal effective volume. Curves for normal (solid) and inverted (dashed) hierarchies are shown. The lines indicate the cut value, where 99% of all supernovae are retained.

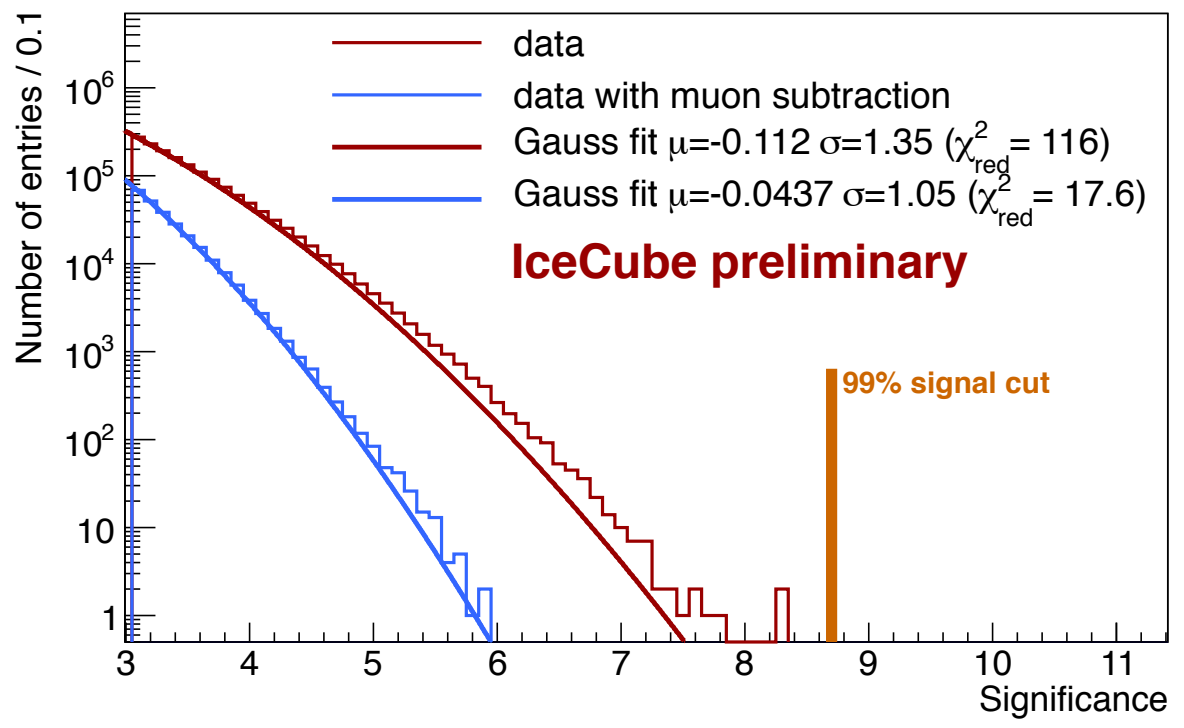

**Figure 5**: Closeup of the data near the cut value for the Hüdepohl model. The muon subtracted data (blue) is kept blind for significances $> 6\sigma$.

Fig. 5 shows the significance of the data distribution with (blinded for $\xi > 6\sigma$) and without muon subtraction close to the cut value. The un-blinded results will be presented at the conference. Assuming that no signal is seen, an upper limit will be provided that is valid for 99 % of all galactic core collapse supernovae with neutrino fluxes equal or higher than in the conservative 8.8 $M_\odot$ Hüdepohl model at the 90 % confidence level. For this model, this corresponds to a cut at $\xi > 8.7\sigma$.

## 4 Conclusion

We set up a search for neutrinos from core collapse supernovae in our galaxy using IceCube data taken between April 2008 to May 2011. The result of this search, valid under conservative assumptions on the distribution of progenitors in the Milky way, their mass and type and accounting for systematic uncertainties, will be presented once the additional two additional years of data taking have been included. The analysis will also be further improved by extending the search to the Magellanic Clouds and employing improved noise reduction techniques.

# Measurement of neutrino oscillations with the full IceCube detector

THE ICECUBE COLLABORATION[1].
[1]*See special section in these proceedings*
*juan.pablo.yanez@desy.de*

**Abstract:** We present preliminary results of the measurement of neutrino oscillations using the first year of data of the completed IceCube Neutrino Observatory. The DeepCore subarray is used to record atmospheric neutrinos that cross the Earth with energies as low as 10 GeV. The IceCube detector is employed to veto the background of muons produced by cosmic rays interacting in the atmosphere. The study benefits from tools designed to diminish the impact of systematic uncertainties and reliably reconstruct neutrinos at the detector's energy threshold. In 343 days of livetime we find 1487 neutrino events. An analysis is performed on the shape of the two-dimensional energy-zenith angle distribution and, in the two flavor approximation, the oscillation parameters that best describe the data are $\sin^2(2\theta_{23}) = 1\,(>0.93)$ and $|\Delta m^2_{23}| = 2.4 \pm 0.4 \cdot 10^{-3}$ eV$^2$.

**Corresponding author:** Juan Pablo Yáñez
*DESY, D-15735 Zeuthen, Germany*

## 1 Introduction

It is a well established phenomenon that neutrinos change their flavor eigenstate during propagation, known as neutrino oscillations. Experiments have tested a wide range of energies and propagation distances finding that a minimal extension to the standard model, with 3 massive neutrinos, can accomodate most measurements made[1].

The value of the parameters that govern this phenomenon have to be obtained from data. For certain experimental set-ups it is possible to assume, with an accuracy of up to a few percent, that only two neutrinos are involved. In such cases, the probability for observing a neutrino with a flavor different from the original is given by

$$P(\nu_\alpha \to \nu_\beta) = \sin^2(2\theta)\sin^2(1.27\Delta m^2 L/E). \quad (1)$$

Here $\theta$ is the mixing angle between the two neutrino mass eigenstates, $\Delta m^2$ is the difference of the square of their masses in eV$^2$, $L$ is the distance traveled in kilometers, and $E$ the neutrino energy in GeV.

This proceeding describes a search for the disappearance of muon neutrinos produced in the atmosphere and detected by the IceCube+DeepCore detector. IceCube is a cubic-kilometer neutrino detector installed in the ice at the geographic South Pole [1] between depths of 1450 m and 2450 m. The DeepCore subarray as used in this analysis includes 7 standard strings plus 8 with denser and more sensitive instrumentation, which results in an energy threshold of $\mathcal{O}(10)$ GeV [2]. The detection and reconstruction of neutrinos relies on the optical detection of Cherenkov radiation emitted by secondary particles produced in neutrino interactions in the surrounding ice.

Atmospheric neutrinos arise primarily from the decay of pions and kaons produced after interactions of cosmic rays with the atmosphere. They cover energies from a few MeV up to few hundred TeV, following a steeply falling spectrum [3]. The flavors produced, in order of abundance, are $\nu_\mu$, $\overline{\nu}_\mu$, $\nu_e$ and $\overline{\nu}_e$.

## 2 Analysis strategy

The current knowledge of oscillations indicates that maximum disappearance is expected for muon neutrinos of about 24 GeV after propagating 12,700 km, roughly the diameter of the Earth. For higher energies the disappaearance effect is reduced, being less than 5 % at 200 GeV. For lower energies the maximum is reached in shorter distances where the incoming direction is shifted towards the horizon.

Muon neutrinos are detected after they interact via charged current (CC) deep inelastic scattering. In a typical $\nu_\mu$ interaction at $E_\nu = 24$ GeV about half of the neutrino energy is transferred to the outgoing muon and half to the shower of hadrons at the interaction vertex. A 12 GeV muon in ice has a range of about 60 m, its primary energy loss mechanism being ionization. The shower has negligible elongation but is considerably brighter, with the hadrons depositing most of its energy as photons. The DeepCore subarray has string-to-string spacings from 40 m to 70 m in the plane parallel to the surface, and 7 m between photosensors in the vertical direction. The Cherenkov light from the detectable secondaries that constitute our signal can reach on average 15 digital optical modules (DOMs) distributed on up to 4 strings. The signal is so faint that the incomplete knowledge of the detector and the medium can have a large impact on its interpretation. These effects were found and reported in a previous analysis that searched for oscillation effects with a partial detector configuration [5].

In this work, complementary techniques for reducing the background are implemented, which are discussed in Section 3. We introduce a method designed to reduce the influence of the medium details on the outcome of reconstruction, described in Section 4. Remaining systematic uncertainties are parametrized and included in the estimation of the oscillation paramters, as discussed in Section 5.

1. For a review of neutrino oscillations see chapter 13 in [4].

## 3 Background reduction

There are two kinds of background events that affect this search. The first and largest are atmoshperic muons that trigger the detector at a rate $10^6$ higher than the neutrinos. The second comes from the neutrinos themselves where neutral current interactions from all neutrino flavors, as well as CC interactions from $\nu_e$, make the oscillation signature weaker.

The atmospheric muon background is reduced by selecting events that start inside the fiducial volume. The DeepCore fiducial volume is situated at the bottom half of the IceCube detector and is surrounded by three layers of standard strings, which define the veto volume for the analysis. Different algorithms are used to search for hits in the veto region that could be connected with the event that triggered DeepCore. They do so by analyzing the event's topology: the position of the first hit of the trigger, the distribution of hits as a function of time and the existence of clusters of causally connected hits in the veto region and isolated hits along directions with sparse instrumentation. After applying these cuts the signal is reduced by half providing a signal to background ratio of 2:1.

Quality cuts are applied on the result of the directional fit, explained in the following section. Functions for the expectation of a track and cascade are fit; the $\chi^2_{cascade}/\chi^2_{track}$ ratio is used to remove all-flavor neutral currents and CC events where only photons from the cascade at the interaction point are seen. A loose cut on the reduced $\chi^2$ of the reconstruction is also applied to further reduce the number of poorly reconstructed events.

## 4 Selection and reconstruction of signal events

### 4.1 Event selection

When a Cherenkov light cone intersects a string, the arrival time of the photons at given depths may be described by a hyperbola. The exact shape depends on the orientation of the cone and the distance between the string and the cone's axis. Figure 1 depicts the pattern expected for two different cone orientations; photons arriving without a significant time delay produce a pattern of hits with these unambiguous signatures.

The first step of the selection is the search for events that contain clusters of hits in hyperbolic patterns. The DOM with the largest photon count serves as a seed and from it a scan is performed to determine if adjacent modules are included. The criterion is entirely defined by causality conditions with respect to the surrounding modules accepted, and the typical jitter and noise rates.

For a given event, each string is independently scanned. A minimum of three accepted DOMs are necessary in order to form a cluster. Events are selected if they contain at least five accepted DOMs. This selection is the most stringent condition applied but assures good event quality and, since it focuses on photons without time delay, removes the need for detailed knowledge of scattering parameters in the ice.

### 4.2 Event reconstruction

The distance that an atmospheric neutrino has traveled to the detector depends on its incoming zenith angle, reconstructed using the previously mentioned clusters of hits. Due to the cluster selection, the fit may be performed without the need of assuming any scattering of the photons. The Cherenkov cone expectation is fit following the method presented in [6]. It is worthwhile to note that by means of using this procedure it is possible to obtain reliable zenith angle reconstructions for events with clusters of hits in only one string. This decreases the acceptance energy threshold of the analysis for up-going directions to about 7 GeV where the disappearance signal is strongest.

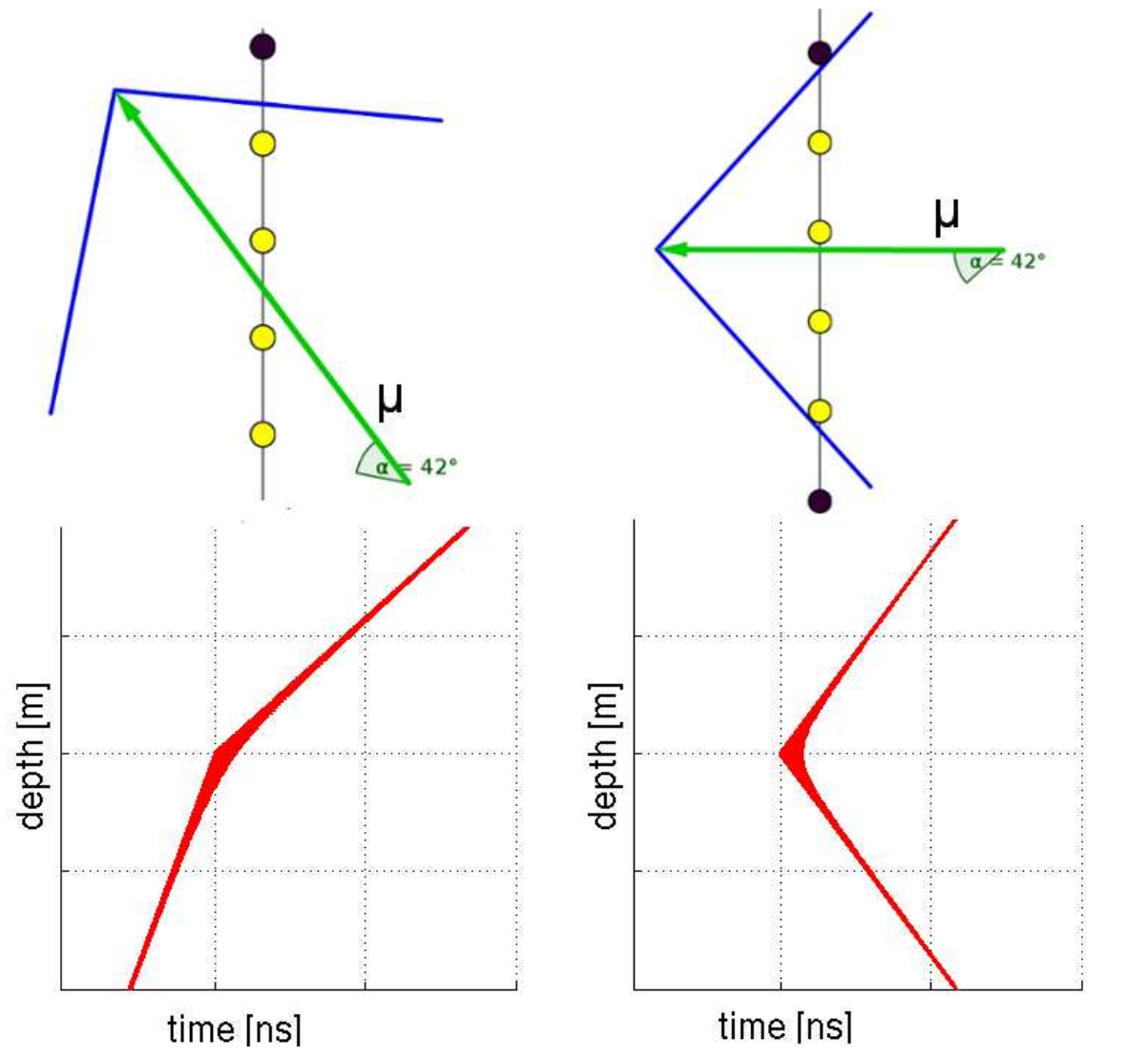

**Fig. 1**: Hyperbolic patters created by the projection of the Cherenkov cone light of a muon track in one string for two different zenith angles.

The total energy of the neutrino is estimated from the light deposition of the hadronic cascade at the vertex and the length of the muon track. Using a fixed direction from the previously described angular reconstruction, an algorithm searches for the first point where a cascade can be accomodated. Following the direction of the event, a subset of hits is extracted from which the average light expectation of a muon track has been removed. This subset is assumed to come from the hadronic cascade and the most likely cascade energy to produce this hit pattern is deduced. The fit algorithm then moves along the direction to search for the best fit point of the end of a muon track with pure Cherenkov emission (other processes, like stochastic energy losses, are not considered). Muon tracks at the energies under discussion here have a roughly constant energy loss of 0.2 GeV/m, so the total energy of the neutrino can be estimated from the cascade energy ($E_{cascade}$) and the track length ($R_\mu$) as $E_\nu = E_{cascade} + R_\mu/5\,\frac{\mathrm{m}}{\mathrm{GeV}}$.

The energy distribution obtained for the final sample is shown in Fig. 2 after selection and reconstruction steps were applied. The sample, CC $\nu_\mu$ being the dominant component with close to 1/3 disappearing due to oscillations, peaks at an energy of 12 GeV.

Figure 3 shows the performance of the reconstructions used for the final event selection. A zenith angle resoultion of 7 degrees, comparable with the kinematic angle between the neutrino and the muon in the sample, is obtained. The energy, resolved with an error of 0.25 in $\log_{10}$(E), is also shown in Fig. 3 (bottom).

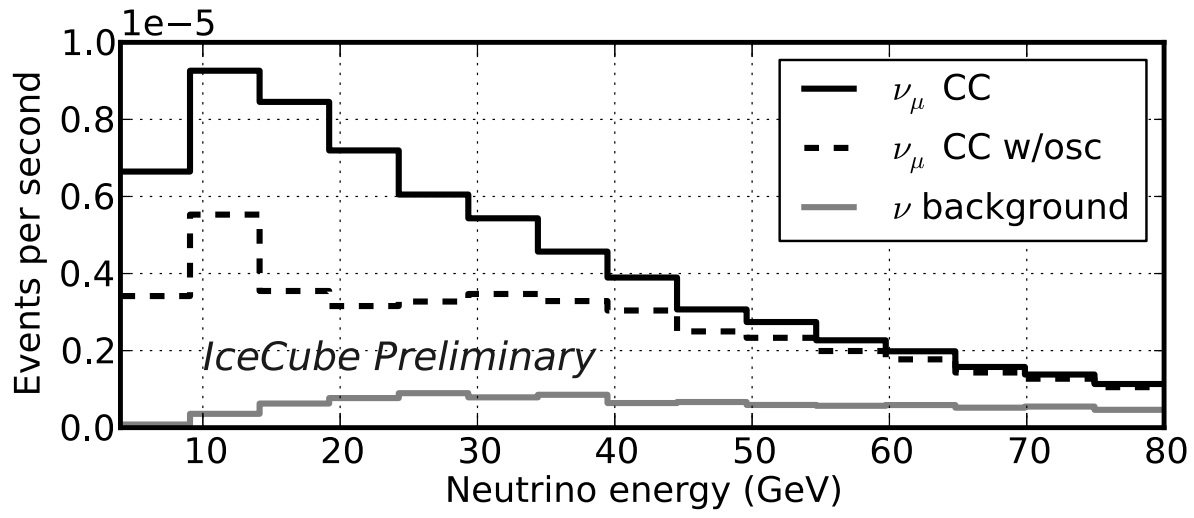

**Fig. 2**: Expected energy distribution for the final neutrino sample. $\nu_\mu$ CC without oscillations (solid black), with oscillations (dashed black) and other $\nu$ contributions (gray).

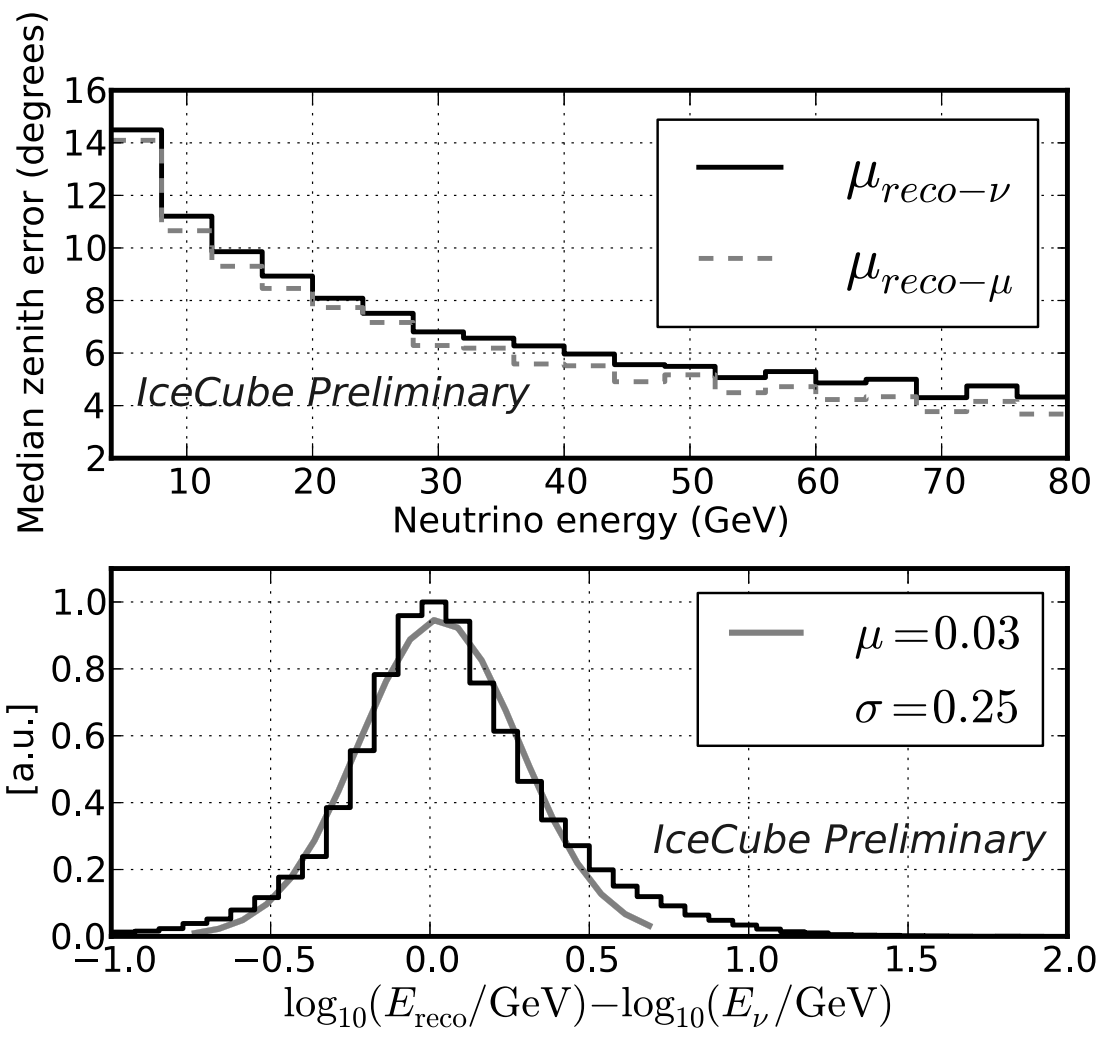

**Fig. 3**: Performance of reconstructions for the $\nu_\mu$ CC component of the final sample . Top: median zenith angle resolution with respect to the neutrino (dashed) and muon (solid) directions as a function of neutrino energy. Bottom: error on the neutrino energy estimation.

## 5 Oscillation parameters extraction

The data were fit by the method of likelihood inference in the presence of nuisance parameters [7]. The observables were binned in an $8 \times 8$ histogram in $\cos(\theta)$ and $\log_{10}(E/[\mathrm{GeV}])$ ranging from [-1, 0] and [0.9, 2] respectively.

### 5.1 Systematic uncertainties

The systematic uncertainties related to the flux of neutrinos include the overall normalization of the flux, the deviation in rate of $\nu_e$ with respect to $\nu_\mu$, and the spectral index. From the detector side the overall optical efficiency and the relative efficiency between DeepCore and IceCube DOMs are included. The properties of the bulk ice and the scattering coefficient $\alpha$ in the refrozen columns of ice where the strings are deployed are also varied. From the event selection the relative contribution of atmospheric muons to the final sample is also taken into account.

Systematic uncertainties that can be parametrized as a function of a single variable are included as nuisance parameters and fit together with the oscillation parameters of interest. When there is prior knowledge for a given parameter a term describing it is added to the likelihood, which has the form

| Nuisance parameter | Prior |
|---|---|
| Atm. $\mu$ contamination | up to 10% |
| Atm. $\nu$ flux | None |
| $\nu_e$ deviation | $\sigma = 20\%$ |
| Spectral index | $\mu = 2.65, \sigma = 0.05$ |
| Optical efficiency | $\sigma = 10\%$ |
| Relative DOM eff. | $\mu = 35\%, \sigma = 3\%$ |
| Ice columns $\alpha$ [1/cm] | $\mu = 0.02, \sigma = 0.01$ |
| Bulk ice properties | See [8] |

**Table 1**: List of systematic uncertainties included as nuisance parameters with their corresponding ranges and priors.

$$-LLH = \sum_{i,j} (\mu_{i,j} - x_{i,j} \ln(\mu_{i,j})) + \frac{1}{2} \sum_k \frac{(q_k - \hat{q}_k)^2}{\sigma_{q_k}^2}. \quad (2)$$

Here the subscripts $i$ and $j$ run over the bins of the energy-zenith angle histogram, where the number of observed events $x$ is compared to the expectation $\mu$, which depends on the oscillation parameters $\theta_{23}$, $\Delta m^2_{32}$ and nuisance parameters $q$. The subscript $k$ runs over the list of nuisance parameters for which prior knowledge exists, penalizing deviations from the mean value $\hat{q}$ in units of its uncertainty $\sigma_q$.

Table 1 contains the list of systematic uncertainties that were considered, as well as their priors and allowed ranges. The first four are related to the relative weight of an event with respect to the sample and can be implemented by modifying those quantities. The remaining systematics must be simulated individually and propagated through the analysis chain in order to be included.

The effects of uncertainties derived from simulation with the exception of the bulk ice properties are applied bin-wise in the energy-zenith angle histogram. The change in the expectation $\mu$ is parametrized as a function of the single variable that describes the uncertainty. They are derived independently and applied as multiplicative factors $g_l$, where $l$ runs over the simulation-derived sources of uncertainty, to the baseline simulation. The number of events expected in the $i,j$-th bin is then

$$\mu_{i,j} = \mu_{i,j_{baseline}}(\theta_{23}, \Delta m^2_{32}; q_w) \prod_l g_l(q_l), \quad (3)$$

where $q_w$ denotes the dependence on the nuisance parameters that can be modified in the weights.

By using this parametrization it is possible to vary all the systematic uncertainties at once, which is a basic requirement for a correct minimization. This approach does assume that the effects of each change can be factorized, which is strictly correct only when discussing individual photons. The impact on event counts could be correlated, but this has not yet been included. Tests have been performed on control samples and, thus far, no abnormal effects have been observed.

The uncertainty on the description of the bulk ice is assessed using simulation generated with different ice models and the best fit oscillation parameters as inferred from the data. The simulation is fit and confidence regions are constructed. The deviations of these confidence regions from the baseline simulation are calculated and added in quadrature to the confidence intervals obtained from the data. This can only be done because most of the the impact of the differences in the ice description are negligible and/or absorbed by the analysis chain. The correction that must be applied is small with respect to the effect of other systematic shifts. All results are shown with this correction clearly separated.

## 6 Results and Discussion

The analysis presented was applied to the first year of data taken by IceCube, from May 2011 until April 2012. In 343 days of livetime, 1487 neutrino events were found. A scan of the oscillation parameters was performed, the results of which are summarized in Fig. 4. The best fit, as well as the regions corresponding to the 68 % and 90 % confidence levels, as calculated from the LLH ratio, are shown. Two sets of confidence contours appear in Fig. 4 corresponding to those with all systematic uncertainties included (gray) and those without the uncertainties on the ice description (black).

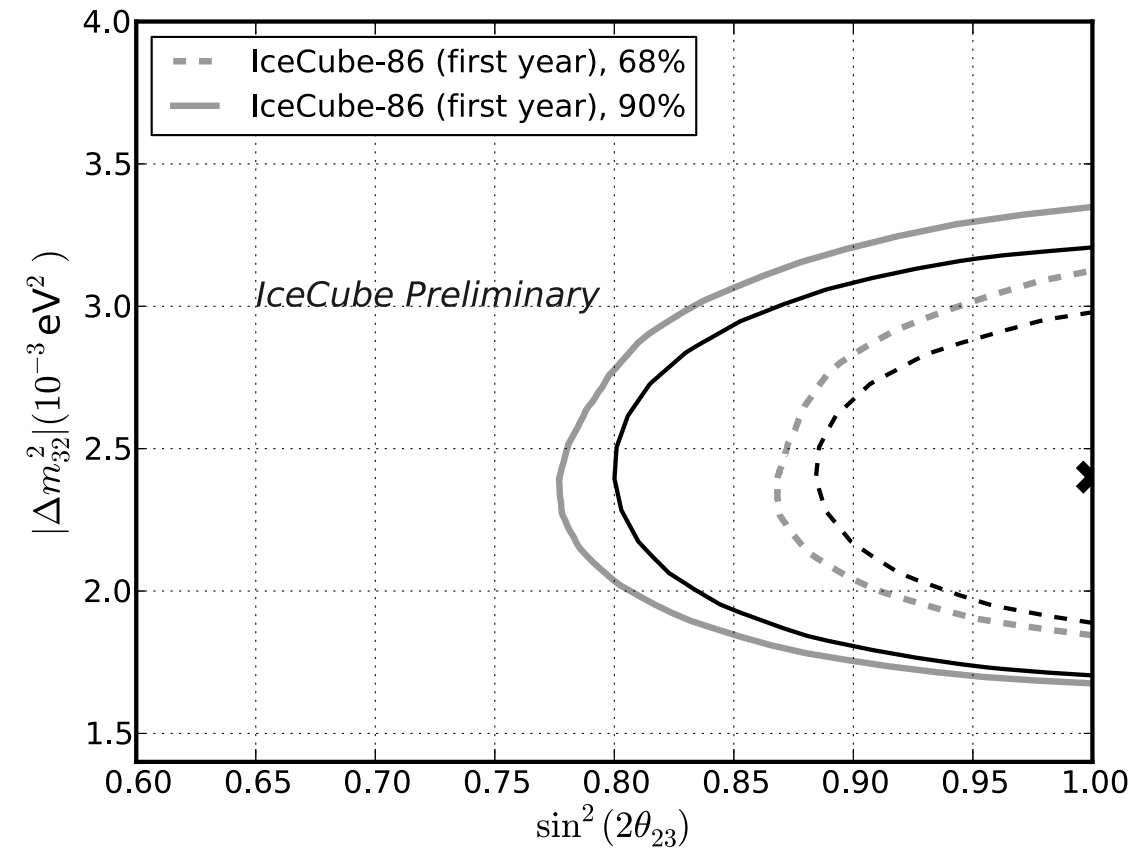

**Fig. 4**: Best fit, 68 % and 90 % confidence intervals. Contours including all systematic uncertainties (gray) and neglecting the uncertainties on the description of the bulk ice (black).

The best fit values for single parameters obtained from the likelihood profile are $\sin^2(2\theta_{23}) = 1\,(> 0.93$ at 68 % C.L.) and $|\Delta m^2_{32}| = 2.4 \pm 0.4 \cdot 10^{-3}$ eV$^2$. Table 2 contains the values of the nuisance parameters at the best fit point.

The agreement between data and simulation was tested using a $\chi^2$. The result is 48.8/54 d.o.f. The energy and zenith angle distributions of the sample and the best fit simulation, which includes the fit nuisance parameters, are shown in Fig. 5.

## 7 Summary and Conclusions

A method to measure neutrino oscillations with the full IceCube detector has been described. It uses reconstruction techniques that aim to reduce the impact of systematic uncertainties of the medium and implements the remaining uncertainties as nuisance parameters to be constrained by the data. Preliminary results are presented that show improvement with respect to a previous measurement [5]. Results will be updated when more data are available.

| Nuisance parameter | Best fit |
|---|---|
| Atm. $\mu$ contamination | 7.6 % |
| $\nu_e$ deviation | -0.5 % (0.02 $\sigma$) |
| Spectral index | 2.66 (0.2 $\sigma$) |
| Optical efficiency | +2.7 % (0.3 $\sigma$) |
| Relative DOM eff. | 35.1 % (0.05 $\sigma$) |
| Ice columns $\alpha$ [1/cm] | 0.018 (0.15 $\sigma$) |

**Table 2**: Values (and deviations in units of sigma) for the nuisance parameters at the best fit point between data and simulation.

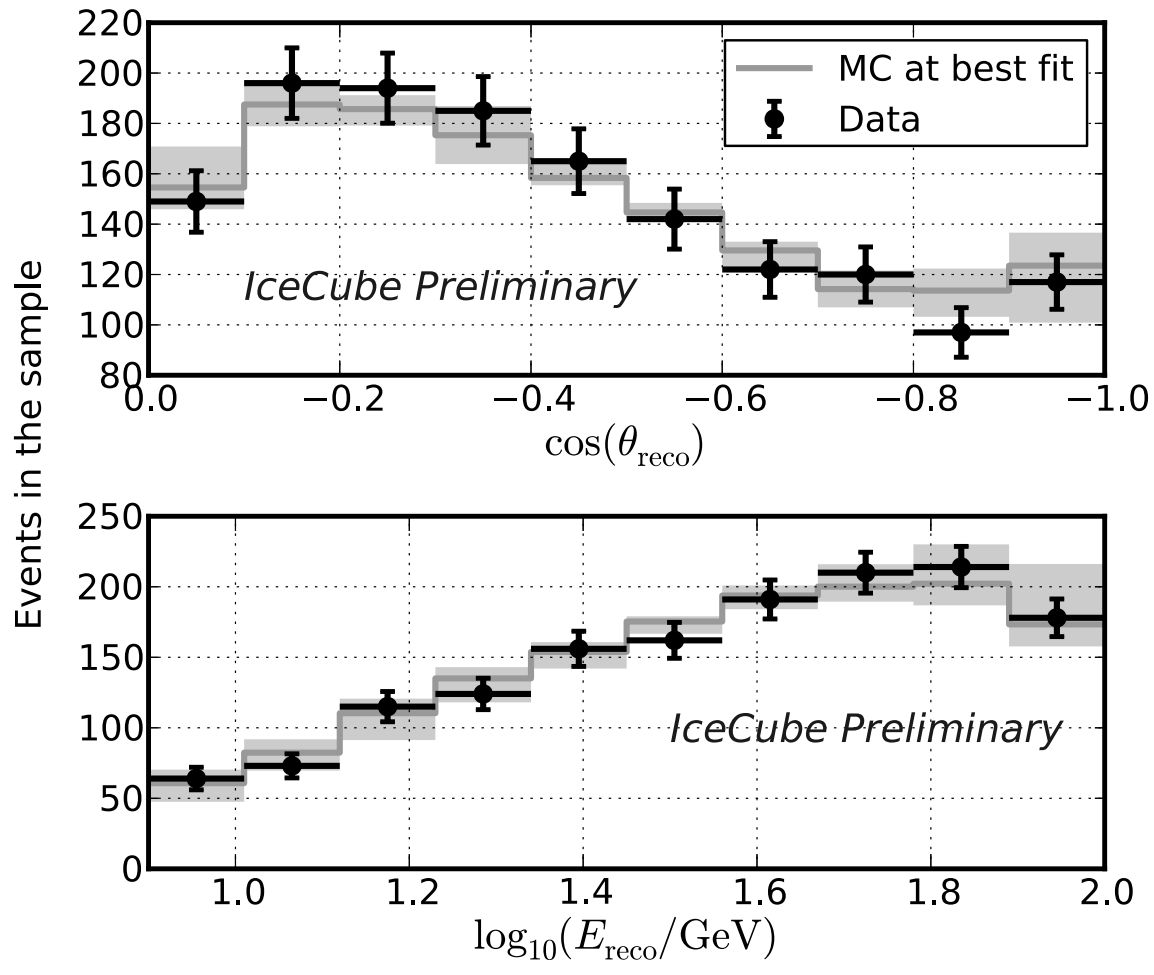

**Fig. 5**: Zenith angle and energy distributions of data and best fit simulation. Statistical errors are added to the data; systematic errors are attributed to the simulation.

# Search for sterile neutrinos with the IceCube Neutrino Observatory

THE ICECUBE COLLABORATION[1],

[1] *See special section in these proceedings*

*marius.wallraff@physik.rwth-aachen.de*

**Abstract:** The IceCube Neutrino Observatory is a 1 km$^3$ Cherenkov detector located at the geographic South Pole. It records atmospheric muon neutrinos with unprecedented statistics of several tens of thousands of identified neutrino events per year and has proven to be suitable for the measurement of muon neutrino disappearance due to neutrino oscillations. Similarly, IceCube is able to search for additional states of sterile neutrinos with mass differences on the order of 1 eV.
If additional sterile neutrino states exist, they will cause unique disappearance signatures for muon neutrinos in the energy range of a few TeV due to matter effects. The survival probability depends on the energy and the path of the neutrino through the Earth and thus its zenith angle. The high statistics and resolutions in the relevant range of energies and baselines make IceCube an ideal tool for testing models of one or more sterile neutrinos.
This work will present an analysis that investigates this signature, using one year of data taken with the IceCube 59-string configuration. It will also discuss the sensitivity that can be reached with five years of data, taken by the 86-string configuration.

**Corresponding author:** M. Wallraff[1]
[1] *III. Physikalisches Institut, RWTH Aachen University*

## 1 Introduction

Neutrino oscillations have become a large research topic over the last decades. The fact that neutrinos are not massless is well established and is the first obvious deviation from the current Standard Model of Particle Physics[1].

Recent results from different areas of neutrino oscillation experiments indicate that there might be effects that can not be explained well with the current model of neutrino oscillations. Examples for such results are the $\bar{\nu}_\mu \to \bar{\nu}_e$ measurements of the two neutrino beam experiments LSND[2] and MiniBooNE[3], as well as various reactor neutrino rate measurements, which see fewer electron antineutrino events than expected within the current model[4]. One way to explain these results is to add additional flavors of neutrinos in a mass range of $\Delta m^2_{42} \approx 1\,\mathrm{eV}^2$. These additional flavor states do not participate in the weak interaction, hence they are called sterile[5]. Because of this property, their existence would not violate the well-known limit on the number of light neutrinos from $Z_0$ branching ratio measurements at the LEP[6].

## 2 The IceCube Neutrino Observatory

The IceCube Neutrino Observatory is a neutrino detector located at the Amundsen-Scott South Pole Station. It uses the naturally clear ice of Antarctica as optical medium to observe Cherenkov radiation emitted by charged leptons that have been created by neutrino interactions in the ice. Its active volume of about one cubic kilometer lies at a depth of about 1.5 to 2.5 km beneath the surface of the ice and is instrumented with 5160 digital optical modules (DOMs) evenly distributed over 86 vertical cables called strings. IceCube was completed in December 2010, but it has already been taking data in the years before. The analysis presented here is conducted on data taken with the 59-string configuration between May 2009 and May 2010, called IC59.

Due to its large size, the IceCube detector is triggered by on the order of 100 000 neutrinos per year. However, because of the small interaction probability of neutrinos with matter, the rate with which muons generated in cosmic-ray air showers (i.e., atmospheric muons) trigger the detector is higher by many orders of magnitude. Atmospheric muons can reach IceCube only from above because they can not traverse more than a few kilometers of matter, so tracks that are upward-going are a good indication for neutrino-induced muons. Unfortunately, a small fraction of the atmospheric muon tracks is misreconstructed as upwards-going and – because of their much higher rates – dominates the triggered dataset.
To clean the dataset of atmospheric muons and badly reconstructed tracks, it is necessary to restrict it to high-quality events. The event selection used for this analysis yields approximately 22 000 muon neutrino events, contaminated by less than 70 atmospheric muon events and is discussed in detail in [7], where it is used in a search for extragalactic high-energy muon neutrinos.

## 3 Neutrino Oscillations in Matter

The analysis discussed here is a neutrino disappearance analysis, conducted with muon neutrinos that are created in cosmic-ray air showers in the Earth's atmosphere. To illustrate the basic principle, it is helpful to first regard the simple formula for the probability of a flavor change for two-flavor oscillations in vacuum,

$$P = \sin^2(2\theta_{\mathrm{mix}})\sin^2\left(1.267\frac{\Delta m^2 L}{E_\nu}\frac{\mathrm{GeV}}{\mathrm{eV}^2\,\mathrm{km}}\right).$$

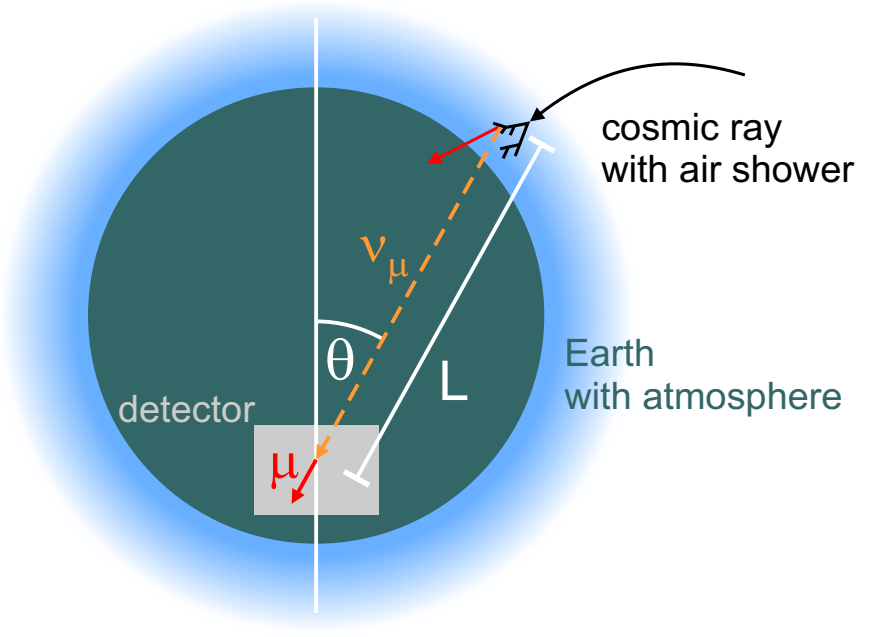

**Figure 1**: Sketch of the Earth with the IceCube detector and a cosmic-ray shower with atmospheric neutrino to illustrate the zenith angle $\theta$ and the oscillation path length $L$.

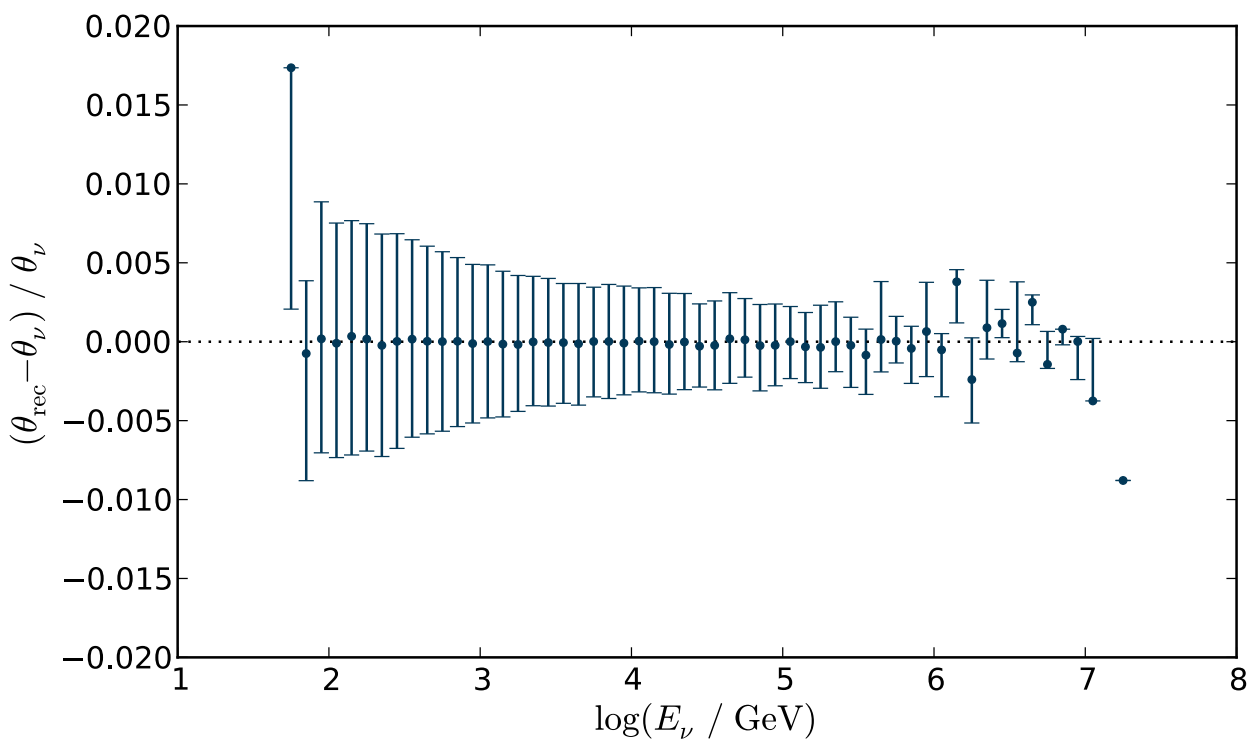

**Figure 2**: Resolution of the reconstructed zenith angle $\theta$ for a simulated dataset of the final event selection. Shown is the median and the 25% and 75% quantile of $\theta$ in dependence of the true neutrino energy $E_\nu$.

$\theta_{\mathrm{mix}}$ and $\Delta m^2$ are parameters that are explained below, $E_\nu$ is the neutrino energy, and $L$ is the length the neutrino traveled after it was created in an interaction, also called baseline. To experimentally determine the constants $\theta_{\mathrm{mix}}$ and $\Delta m^2$, one can compare the rate of measured neutrinos with the rate of neutrinos expected from simulations at various values of $L$ and $E_\nu$ and find the best matching values for the constants. As IceCube measures atmospheric neutrinos that travel up to two Earth radii before they reach the detector, it has access to baselines between about 150 km and 12 000 km, and to neutrino energies between about 10 GeV and 100 TeV. The baseline is defined by the zenith angle $\theta$ of the neutrino as can be seen in fig. 1. For the energies above 100 GeV that are relevant for this analysis, this angle can be measured with great precision and accuracy, see fig. 2. The neutrino energy can be inferred by measuring the energy of the induced muon. As can be seen in fig. 3, the energy resolution is somewhat limited because of the fact that the induced muon does not get the full energy of the neutrino, and that the interaction might have happened an unknown distance outside the detector. However, as shown below, it is sufficient given the statistical precision that can be achieved with IceCube.

To understand the signatures this analysis is looking for, it is necessary to replace the two-flavor vacuum model by a more complex one. Neutrino oscillations between $N$ flavors of neutrinos occur due to the $N$ mass eigenstates $\nu_i$ not being identical to the $N$ flavor eigenstates regarding the weak interaction, $\nu_\alpha$. Instead, the eigenstates can be described as linear combinations of each other, given by the Pontecorvo-Maki-Nakagawa-Sakata (PMNS) matrix $U$:

$$\nu_\alpha = \sum_i U_{\alpha i} \nu_i$$

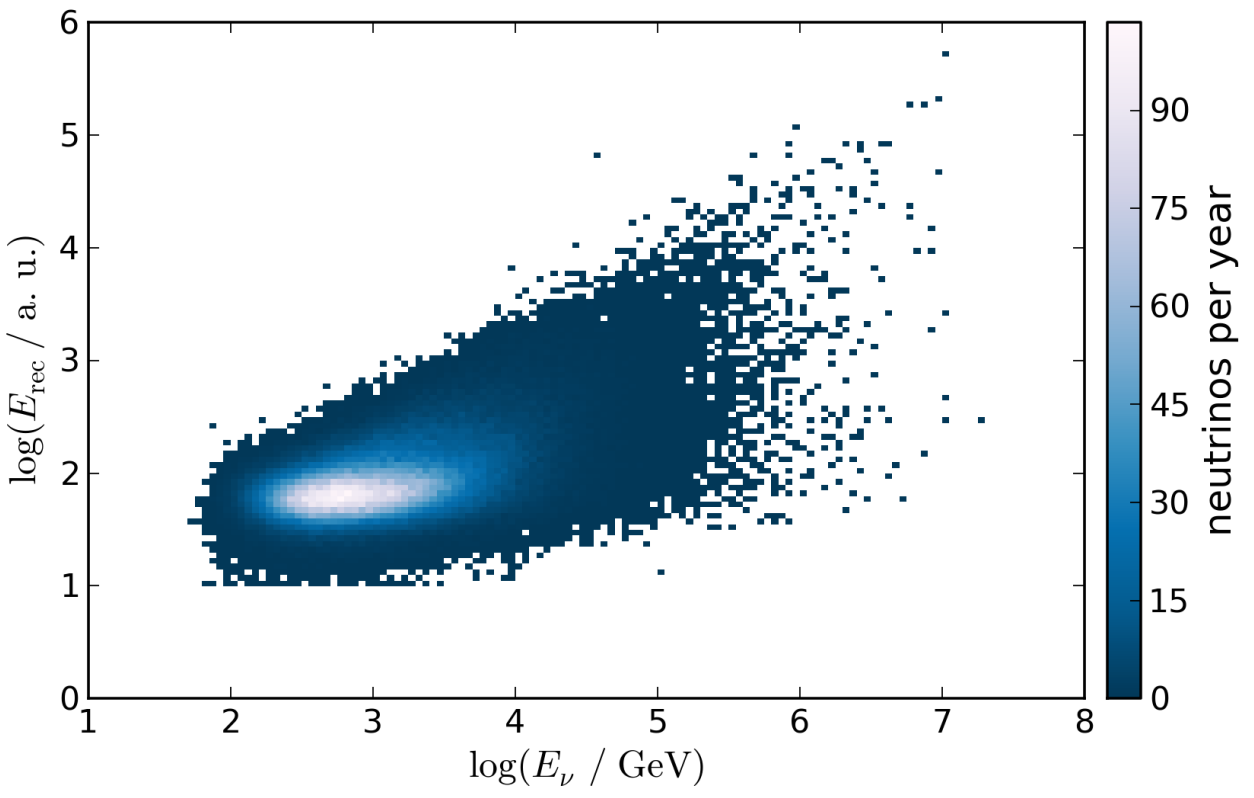

**Figure 3**: Correlation of the reconstructed neutrino energy $E_{\mathrm{rec}}$ with the true neutrino energy $E_\nu$ for a simulated dataset of the final event selection.

$U$ can be parametrized as a product of rotation matrices $R_{ij}(\theta_{ij}, \delta_{ij}) \in \mathbf{C}^{N\times N}$,

$$U = \prod_{i=1}^{N-1} \prod_{j=i+1}^{N} R_{ij},$$

where $\theta_{ij}$ are the mixing angles and $\delta_{ij}$ are CP-violating Dirac phases. In general, $U$ also gets multiplied by a diagonal matrix containing CP-violating Majorana phases, but these do not have any influence on oscillation effects and can therefore be ignored for this work. The parametrization has to be given explicitly because rotation matrices do not commute and for $N > 3$ there is no canonical parametrization. In total, $U$ depends on $\frac{1}{2}N(N-1)$ mixing angles, but only on $\frac{1}{2}(N-1)(N-2)$ Dirac phases because additional phases are redundant.

Using the PMNS matrix $U$, the oscillation probabilities can then be computed numerically by solving the $N$-dimensional Schrödinger equation

$$i\frac{\mathrm{d}}{\mathrm{d}x}\left|\nu_{\mathrm{flavor}}\right\rangle = \frac{1}{2E_\nu}(H_0 + A)\left|\nu_{\mathrm{flavor}}\right\rangle,$$

where $H_0 = U \cdot \mathrm{diag}(\Delta m_{i1}^2) \cdot U^\dagger$ depends on the squared neutrino mass differences $\Delta m_{i1}^2 = m_i^2 - m_1^2$. The choice of $m_1^2$ as point of reference is arbitrary, as every matrix proportional to identity does not influence the oscillation probabilities. $A$ describes an effective squared mass, induced by the matter the neutrinos traverse. For this induced mass, only interactions are relevant in which the neutrino is preserved: The three conventional flavors can scatter on nucleons and electrons by neutral-current (NC) interactions, and electron neutrinos can additionally scatter by charged-current (CC) interactions on the many electrons present in Earth's matter without the neutrino being absorbed. Analyses that investigate the first oscillation maximum between muon and tau neutrinos can typically neglect matter effects because muon and tau neutrinos see the same matter potential. However, sterile neutrinos lack possibilities to interact weakly by definition, which leads to strong matter effects in the oscillation between sterile and conventional neutrinos. The effective squared mass term becomes $A = \mathrm{diag}\,(A_{\mathrm{CC}} + A_{\mathrm{NC}}, A_{\mathrm{NC}}, A_{\mathrm{NC}}, 0, \ldots)$ with $A_{\mathrm{CC}} = \pm 2\sqrt{2} G_{\mathrm{F}} \rho N_{\mathrm{A}} Y_{\mathrm{e}} E_\nu$ and $A_{\mathrm{NC}} = \mp\sqrt{2} G_{\mathrm{F}} \rho N_{\mathrm{A}} Y_{\mathrm{n}} E_\nu$, where $G_{\mathrm{F}}$ is Fermi's constant, $\rho$ is the mass density, $N_{\mathrm{A}}$ Avogadro's constant, and $Y_{\mathrm{e}}$ is the electron fraction ($Y_{\mathrm{e}} = Y_{\mathrm{p}} = 1 - Y_{\mathrm{n}}$)[8]. Due to the density-dependent terms, neutrinos can experience oscillation resonances in matter that can strongly enhance oscillation probabilities even for otherwise very small mixing angles[9]. This effect is exploited for this muon neutrino disappearance analysis, significantly improving its sensitivity. The signs of $A_{\mathrm{CC}}$ and $A_{\mathrm{NC}}$ are different for particles and antiparticles and depend on both the conventional mass hierarchy, i.e., $\mathrm{sgn}(\Delta m^2_{32})$, and on the sign of $\Delta m^2_{42}$. Therefore, in general, resonances that enhance oscillation effects for particles suppress them for antiparticles and vice versa. However, IceCube can not distinguish between particles and antiparticles, so for the purpose of this analysis, the probabilities are averaged between both.

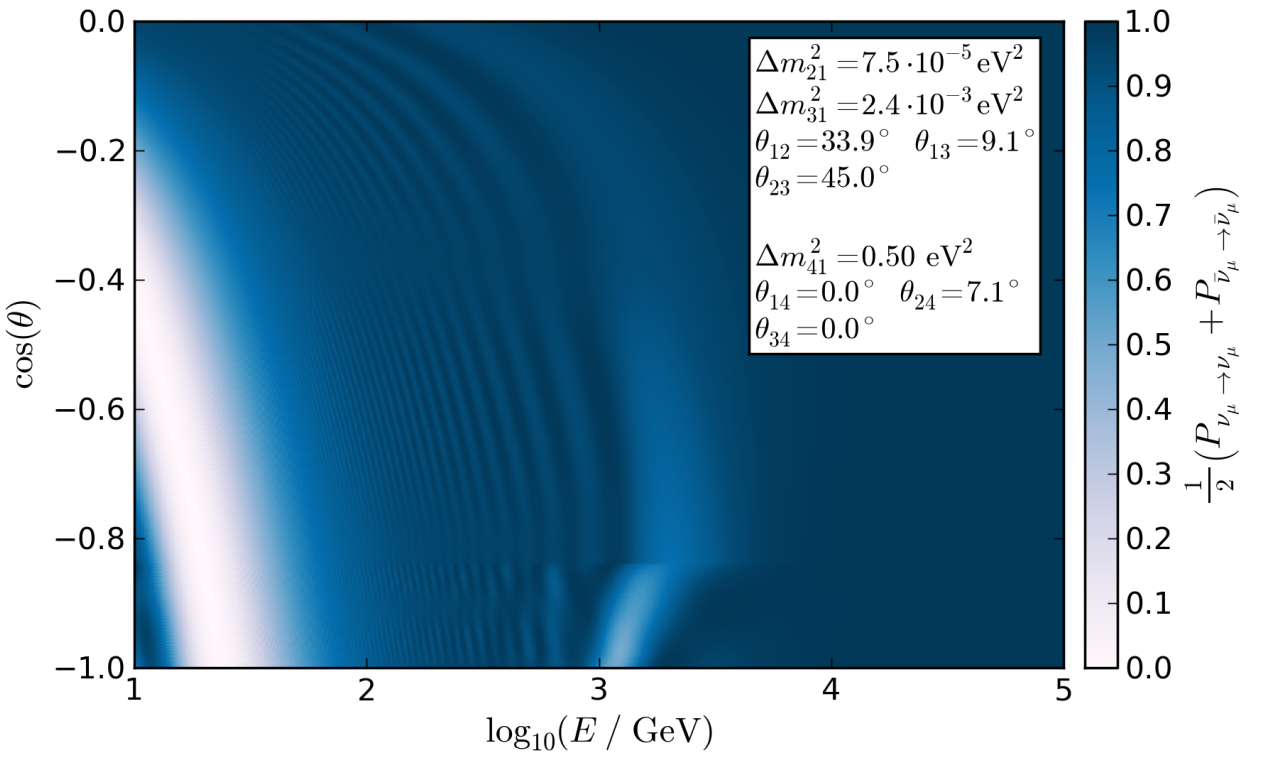

**Figure 4**: Probability averaged over atmospheric muon neutrinos and antineutrinos for the particle to reach the detector in the same state, with one sterile state of neutrinos, parameters as specified in the plot. Does not include detector resolution or acceptance.

## 4 Analysis Method

As described above, the oscillation probability depends on the path the neutrino has taken through the Earth and on its energy $E_\nu$. As the Earth can be approximated as a symmetrical sphere, the path of the neutrino only depends on the zenith angle $\theta$ (fig. 1). The general analysis strategy is to employ a two-dimensional likelihood ratio test to check for the disappearance signature in the atmospheric muon neutrino distribution, binned in the reconstructed values for $\cos(\theta)$ and $E_{\mathrm{rec}}$. For the matter density $\rho$ of the Earth, the Preliminary Reference Earth Model (PREM) is used[10]. The oscillation probabilities have been calculated using the Python tool nuCraft.[1] For details such as the handling of the Earth's atmosphere, please refer to [11].

An example for the atmospheric muon neutrino disappearance signature given one sterile neutrino can be seen in fig. 4. The parameters of the sterile neutrino have been chosen to be $\Delta m^2_{42} = 0.5\,\mathrm{eV}^2$, $\theta_{24} = 7^\circ$, and $\theta_{14} = \theta_{34} = 0^\circ$, because these values have not been excluded by MINOS in 2011[12], yet are large enough to serve well for illustrative purposes. The plot shows the survival probability averaged between muon neutrinos and antineutrinos. A strong oscillation minimum due to a resonance can be seen at about 4 TeV for the particles that traversed the inner core of the Earth. Variation of $\Delta m^2_{42}$ leads to a shift of the minimum on the $E_\nu$ axis, while variation of $\theta_{24}$ shifts the minimum along the zenith angle axis and also steers the depth and shape of the minimum. $\theta_{14}$ controls the mixing between electron and sterile neutrinos and is negligible for the atmospheric muon neutrino disappearance as seen by IceCube, whereas the tau-sterile angle $\theta_{34}$ influences depth and shape roughly similar to $\theta_{24}$, as long as $\theta_{24} \neq 0$[11]. A side-effect of $\theta_{34} \neq 0$ is that a large fraction of the muon neutrinos will not oscillate to sterile neutrinos, but to tau neutrinos instead. However, at the relevant energies of a few TeV, 85% of the tau neutrinos that interact produce cascade-like signatures and the remaining 15% produce faint muon tracks with large initial cascades, so they are strongly suppressed by the event selection that favors track-like signatures.
For this analysis, it was decided to limit the likelihood scan to the physics parameters $\Delta m^2_{42}$and $\theta_{24}$. These two parameters suffice to reproduce all signatures 3+1 models can assume in IceCube, especially considering IceCube's somewhat limited energy resolution. The analysis still remains sensitive to models with 3+2 or more neutrino flavors, because the signatures of these models can be approximated well as superposition of 3+1 signatures, and are therefore much more similar to 3+1 models than to 3+0 models. After the likelihood scan has been conducted, selected models that have not been considered in the scan will be examined.

Instead of a standard Poisson likelihood that only takes into account statistical uncertainties of the measured data, a formulation is used that also takes into account statistical uncertainties of the simulated reference histograms [13]. Using this likelihood, a scan is performed in the parameter range of $\Delta m^2_{42} = 10^{-2.0}..10^{0.7}\,\mathrm{eV}^2$. The likelihood ratios, i.e., the differences between the logarithmic likelihood values for a given pair of parameters and the null hypothesis, $LLH\left(\Delta m^2_{42}, \theta_{24}\right) - LLH\,(0,0)$, are then used as a test statistic. According to Wilks' theorem, two times the likelihood ratios follow a $\chi^2$ distribution with two degrees of freedom if the null hypothesis is true[14]. Although the preconditions of Wilks' theorem are not clearly fulfilled by this analysis, the applicability of Wilks' theorem has been verified empirically.

## 5 Systematic Uncertainties

Systematic uncertainties play an important role in this analysis, as it relies heavily on the agreement between simulated and measured data. All theoretical uncertainties that can have an effect on the energy or the zenith angle spectrum must be taken into account. In this analysis, this is achieved by parametrizing these effects and including them into the likelihood function as nuisance parameters as described in [15]. The meaning of nuisance parameters is that in contrast to the physics parameters in the likelihood function, the fit results for them will not be considered to be measurements of the physical quantities they describe, because they have not been tested to not be degenerate with other nuisance parameters that may or may not have been implemented in the fit.As required by likelihood ratio tests, nuisance parameters get minimized independently for both

1. available at `nucraft.hepforge.org`

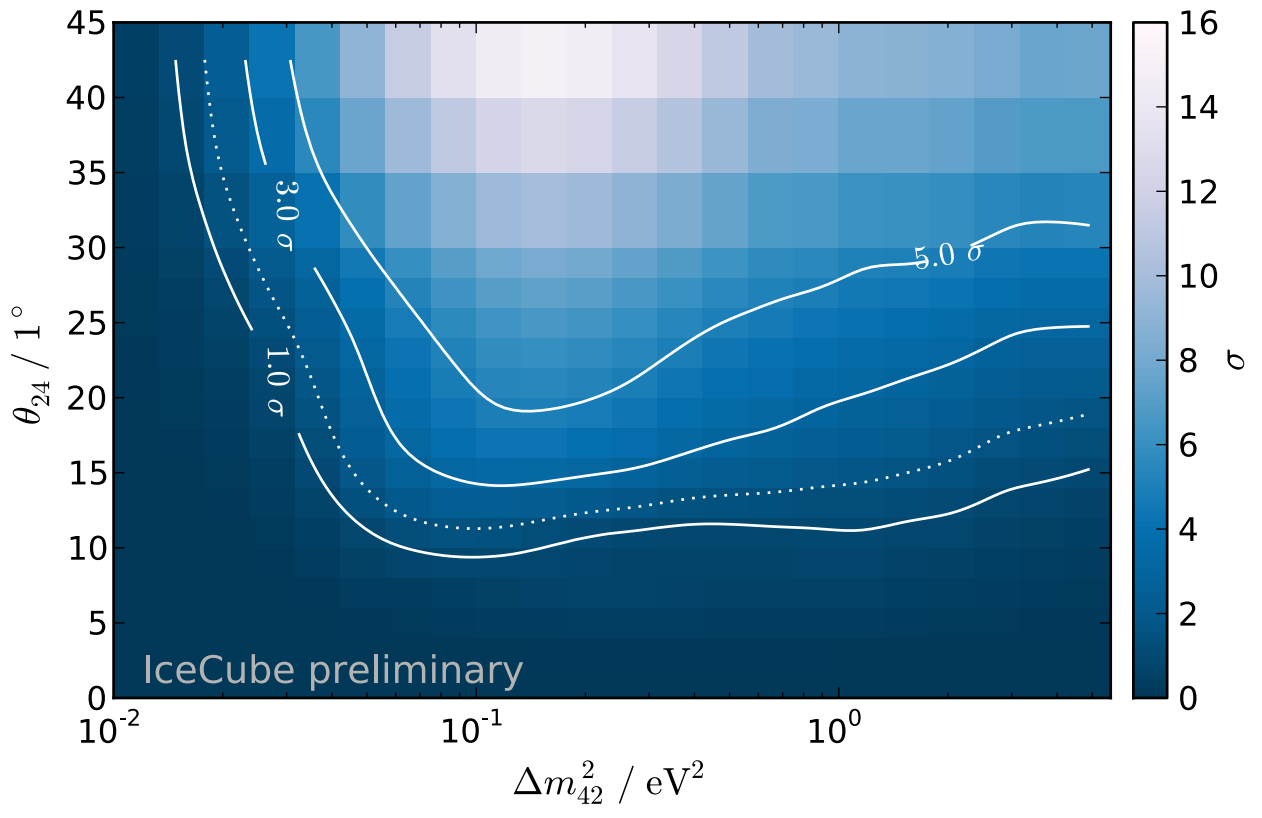

**Figure 5**: Sensitivity for the exclusion of oscillation parameters with the IC59 dataset, given the null hypothesis. The dotted line indicates the 90% confidence level. Systematic uncertainties mentioned in the text have been taken into account (without priors), except for the ice model.

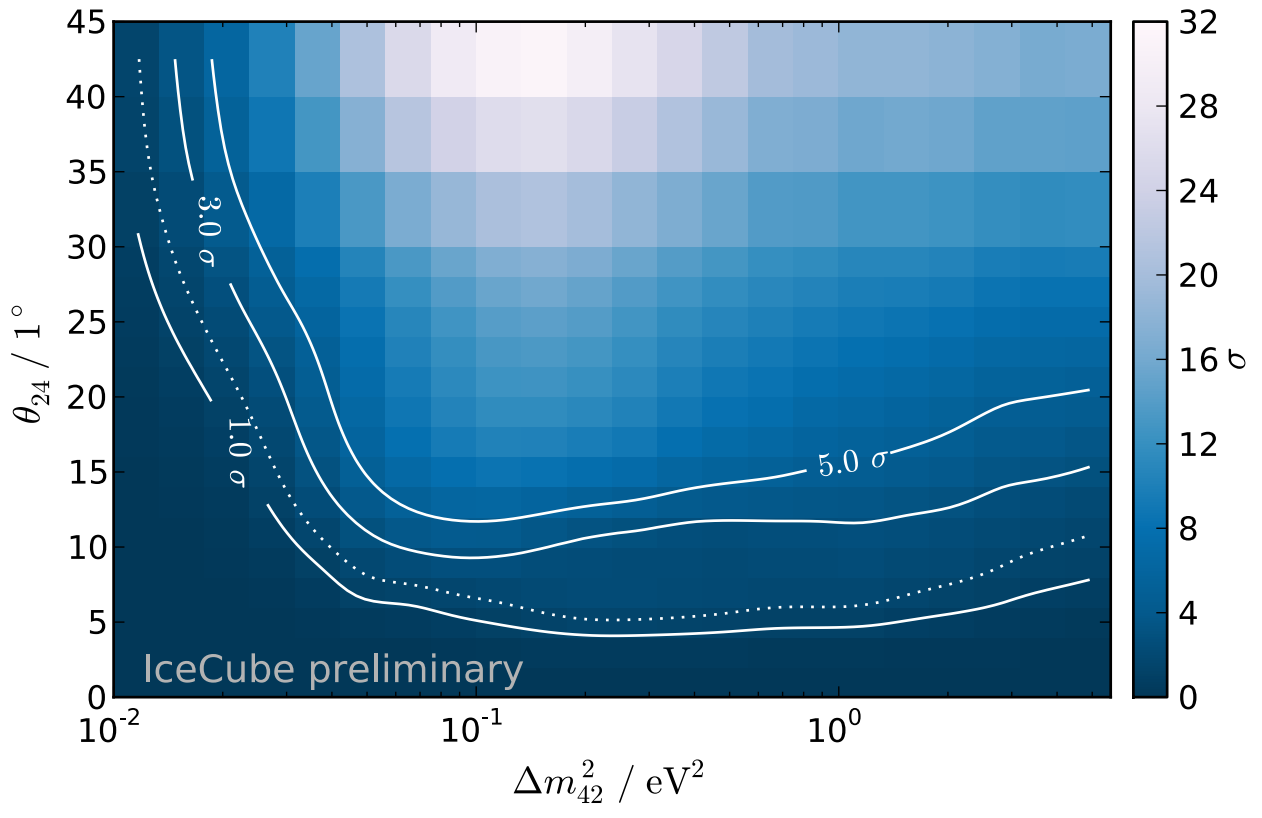

**Figure 6**: Estimated sensitivity for the exclusion of oscillation parameters with five years of IC86 data, given the null hypothesis and the same event selection. The dotted line indicates the 90% confidence level. Systematic uncertainties mentioned in the text have been taken into account (without priors), except for the ice model.

null hypothesis and signal hypothesis at every point of the LLH scan[16].

The continuous nuisance parameters that are included in the likelihood function are the total neutrino flux, the cosmic-ray spectral index, the pion-kaon ratio in cosmic ray interactions, and the relative optical efficiency of IceCube's DOMs. Uncertainties in the optical model of the ice in which IceCube is embedded will be taken into account as a discrete nuisance parameter, which means that for every point in the physics parameter space the best-matching ice model will be chosen. The following plots do not yet include the ice model uncertainty.

## 6 Sensitivity and Outlook

The sensitivity of this analysis has been estimated using the so-called Asimov approach, in which it is derived from the most representative simulated dataset as described in [16]. Figure 5 shows the sensitivity with which IC59 can exclude certain pairs of $\Delta m^2_{42}$ and $\theta_{24}$, given the null hypothesis and the remaining oscillation parameters as in fig. 4. It shows that for $\Delta m^2_{42} \approx 0.1\,\mathrm{eV}^2$, $\theta_{24}$ mixing angles of 12° can be probed at a 90% confidence level. At smaller and larger $\Delta m^2_{42}$, the sensitivity diminishes due to the oscillation minimum moving into regions of lower statistics. It is expected that the sensitivity of the analysis improves significantly when priors on the nuisance parameters get implemented that restrict the uncertainties to parameter regions that have not yet been excluded.

Figure 6 shows a projection of this analysis to a five-year dataset taken with IC86, containing 200 000 muon neutrino events. It can be seen that the analysis on IC59 is currently limited by the statistics of the data and not by systematic uncertainties. The projection is conservative in the sense that the event selection could be modified to include more low-energy events, further boosting the sensitivity for low values of $\Delta m^2_{42}$. This should be fairly easy, because from the 79-string configuration onwards, IceCube's low-energy extension DeepCore is available to reach energies as low as 10 GeV. Also, the larger detector allows for better angular and especially energy reconstruction, which is not taken into account by this projection.

The next steps for this analysis are to finalize systematic studies and then to apply the analysis on the experimental data and calculate limits. Logical next steps are the extension to lower neutrino energies to increase the sensitivity for lower values of $\Delta m^2_{42}$, and to move on to multi-year datasets taken with newer configurations of IceCube as they become available.

# Study of the sensitivity of IceCube/DeepCore to atmospheric neutrino oscillations

THE ICECUBE COLLABORATION[1],

[1]*See special section in these proceedings*

*julia.leute@ph.tum.de*

**Abstract:** The IceCube Neutrino Observatory located at the geographical South Pole is the world's largest neutrino detector. One of the design goals of its low energy extension DeepCore is to increase the sensitivity to atmospheric neutrino oscillations. IceCube/DeepCore has already seen atmospheric neutrino oscillations with high statistical significance. In this work we explore the potential of the detector to constrain competitively the atmospheric mixing parameters and to test the maximal mixing hypothesis. Based on the current performance of our oscillation analyses, IceCube/DeepCore can establish non-maximal mixing ($\theta_{23} \neq 45°$) at the $3\sigma$ level with four years of data for $\sin^2(2\theta_{23}) \leq 0.90$ with full systematic uncertainties at our present level of understanding.

**Corresponding authors:** Julia Leute [1], Andreas Groß [1], Elisa Resconi [1],
[1] *Physik-Department, Technische Universität München, Boltzmannstr. 2, D-85748 Garching, Germany*

## 1 Neutrino Oscillations with IceCube/DeepCore

Interactions of cosmic rays in the atmosphere provide us with electron and muon (anti)neutrinos with a wide energy range and from different zenith angles $\phi_z$ corresponding to different average neutrino propagation lengths. Neutrino oscillations have been observed by many experiments covering different energy ranges, propagation lengths and neutrino channels. Recently, the atmospheric muon neutrino disappearance has been measured with high statistical significance by IceCube/DeepCore [1, 2, 3] in an energy range not explored before.

IceCube is a cubic-kilometer neutrino observatory installed in the ice at the geographic South Pole [4] between depths of 1450 m and 2450 m. Detector construction started in 2005 and finished in 2010. This study is based on the performance of the 79-string configuration, which was taking data in 2010 before completion of the detector. The DeepCore subarray [5] as defined in this study includes six densely instrumented strings optimized for low energies and seven adjacent standard strings. Neutrino energy and zenith angle reconstruction relies on the optical detection of Cherenkov radiation emitted by secondary particles produced in neutrino interactions with the deployed digital optical modules (DOMs).

DeepCore lowers the energy threshold to 10 GeV which allows the observation of the first minimum of the muon neutrino survival probability. This minimum is at $\sim 25$ GeV for vertically upgoing neutrinos ($\phi_z = 180°$) travelling $12.7 \cdot 10^3$ km through the Earth before reaching the detector and shifts progressively to lower energies for $\phi_z < 180°$. The muon neutrino survival probability is calculated in the two flavour formalism:

$$P_{\nu_\mu \to \nu_\mu} = 1 - \sin^2(2\theta_{23}) \sin^2\left(1.27 \frac{\Delta m^2_{32} L}{E} \frac{\text{GeV}}{\text{eV}^2\,\text{km}}\right) \quad (1)$$

with the propagation length $L$, the neutrino energy $E$ and the atmospheric oscillation parameters $\Delta m^2_{32}$ (mass splitting) and $\theta_{23}$ (mixing angle).

In this work, we explore the potential of the detector to constrain competitively the atmospheric mixing parameters and to test the maximal mixing hypothesis in the near future when higher statistics samples and better event reconstruction methods for low energy events will be available.

## 2 Monte Carlo Event Sample

This Monte Carlo (MC) study is based on the low energy event selection from [2], but assumes a ten times higher signal statistics as realized in [1]. The simulated data are binned in a two dimensional histogram with ten $\cos(\phi_z)$ bins ranging from -1 to 0 and five $\log(E/\text{GeV})$ bins ranging from 1 to 2.

The zenith angle is reconstructed using the algorithm described in [6] with a median resolution of 8°. The energy estimator is not based on an existing reconstruction algorithm, but instead uses the true (MC) energy with a Gaussian smearing. Inspired by [7], the width of the Gaussian has a constant term and an energy dependent term: $(\sigma_E/\text{GeV})^2 = 5^2 + (0.2 \cdot E/\text{GeV})^2$.

The constant term is meant to describes the uncertainty on the determination of the muon track length (1 GeV corresponds to 5 m), while the energy dependent term accounts for the fluctuations in the fraction of the energy that is transfered to the muon in the neutrino interaction.

Initial studies have shown that a median energy resolution of $\sim 50\%$ at 10 GeV is achievable, and although the assumed resolution has not yet been reached for the higher energies it is a realistic expectation.

## 3 Background

The background for charged current muon neutrino interactions comes from neutral current interactions of all flavours,

atmospheric muons and interactions of atmospheric electron neutrinos and appearing tau neutrinos.

The atmospheric muon background is reduced by using the outer strings surrounding DeepCore as an active veto [8]. In current analyses [1, 2, 3], the muon background is reduced to below 10% of the total event rate, depending on the event selection. Atmospheric electron neutrinos make up $\sim$ 10% of all events.

For this study, we assumed 10% electron neutrinos and 10% tau neutrinos (relative to the muon neutrino deficit) as background.

## 4 Test Statistic

The comparison of the zenith-energy histograms is performed with the covariance approach [9] with the test statistic

$$\chi^2 = \sum_{m,n=1}^{N} (R_n^{exp} - R_n^{theo})[\sigma_{nm}^2]^{-1}(R_m^{exp} - R_m^{theo}) \quad (2)$$

where $N$ is the number of observables (histogram bins) with their corresponding experimental observations $\{R_n^{exp}\}_{n=1,..,N}$ and their predictions $\{R_n^{theo}\}_{n=1,..,N}$ and $\sigma_{nm}^2$ is the covariance matrix

$$\sigma_{nm}^2 = \delta_{nm} u_n u_m + \sum_{k=1}^{K} c_n^k c_m^k \quad (3)$$

with uncorrelated errors $u_n$ ("statistical uncertainties") and $K$ correlated errors $c_n^k$ ("systematic uncertainties").

The atmospheric oscillation parameters are extracted by varying the oscillation parameter values that go into the predictions of the observables and evaluating the test statistic for all parameter values. This yields best fit parameter values at the $\chi^2$ minimum and confidence regions from the $\chi^2$ surface.

## 5 Systematic Uncertainties

Besides the atmospheric oscillation parameters there are additional experimental and theoretical uncertainties that affect the outcome of the measurement systematically. The uncertainties are propagated in the simulation to the final event selection level to estimate the effect on the observables. The underlying assumption is that the uncertainties propagate linearly and changing the $k$-th uncertainty parameter by $x\sigma$ shifts the $n$-th observable by $xc_n^k$.

Six sources of systematic uncertainties, thereof three of experimental origin and three in theoretical predictions, were included in the study. The uncertainty of the optical efficiency of the photon yield and detection by IceCube DOMs is assumed to be $\pm$10% and the relative sensitivity of the high quantum efficiency DeepCore DOMs compared to standard DOMs is assumed to be $(135 \pm 3)$%. To estimate the effect of the uncertainties on the optical ice properties two different ice models [10, 11] were compared. The uncertainties on the atmospheric neutrino fluxes were investigated by varying the overall flux normalization by $\pm$25%, the cosmic ray spectral index by $\pm$0.05 and comparing two different calculations [12, 13] with different assumptions about the hadronic interactions in the atmosphere. The uncertainty on the neutrino cross sections is assumed to be degenerate with the uncertainties on the atmospheric neutrino fluxes and it is not included explicitly as an additional systematic uncertainty.

This uncertainty ranges, as used in current analyses (e.g. [2]), are conservative estimates since we are working on a better understanding and reduction of our systematic uncertainties.

## 6 Sensitivity Calculations

To estimate the median significance at which we could measure the injected oscillation parameters we used the so called Asimov data set [14], the MC prediction for the assumed true parameter values, as pseudo-data. This pseudo-data was then compared to the MC predictions with varying atmospheric neutrino oscillation parameters.

Confidence regions were calculated with a $\Delta\chi^2$ threshold based on Wilks' theorem [15]. We assume a $\chi^2$ distribution of the test statistic with the number of degrees of freedom corresponding to the number of neutrino oscillation parameters used in the fit. Ensemble tests have been performed to test this assumption. The real distribution of the test statistic lies between one and two degrees of freedom, while our calculations assume two degrees of freedom, therefore the constructed confidence regions have an over-coverage. The main reason for this deviation is the proximity of the boundary of the parameter space at $\sin^2(2\theta_{23}) = 1$ which violates the regularity conditions of Wilks' theorem.

For the sensitivity to the exclusion of maximal mixing as a function of $\sin^2(2\theta_{23})$ we fitted the atmospheric oscillation parameters for various injected $\sin^2(2\theta_{23})$ values and $\Delta m_{32}^2 = 2.4 \cdot 10^{-3}$ eV$^2$. We minimized the test statistic $\chi^2$ over $\Delta m_{32}^2$ to take into account the degeneracy of the atmospheric neutrino oscillation parameters. The median significance is then given by the $\chi^2$ difference between the best fit (which is at the injected parameter value when using the Asimov data set) and $\sin^2(2\theta_{23}) = 1$. The test statistic is assumed to follow a $\chi^2$ distribution with one degree of freedom due to the minimization over $\Delta m_{32}^2$.

Figure 1 shows the sensitivity to the exclusion of maximal mixing depending on $\sin^2(2\theta_{23})$ that can be achieved with higher statistics samples.

IceCube/DeepCore can establish non-maximal mixing at the $3\sigma$ level with four years of data for $\sin^2(2\theta_{23}) \leq 0.90$ with full systematic uncertainties. With six years of data IceCube/DeepCore measurements will also be competitive with current constraints of the atmospheric neutrino oscillation parameters (see Figure 2).

## 7 Discussion

The challenge for future analyses will be to keep more events while maintaining a high quality and purity of the event sample. While the sample used in this study has an energy distribution that peaks at 40 GeV, further improvement is expected from the inclusion of more low energy events down to below 10 GeV, where degeneracies between the atmospheric neutrino oscillation parameters are reduced. At these energies three flavour effects become important, which might give rise to a sensitivity to the octant of the atmospheric mixing angle. In the two flavour approximation the muon neutrino survival probability depends on $\sin^2(2\theta_{23})$ and is therefore symmetric around $\theta_{23} = 45°$, whereas the three flavour formalism distinguishes between $\theta_{23} < 45°$ (lower octant) and $\theta_{23} > 45°$ (higher octant).

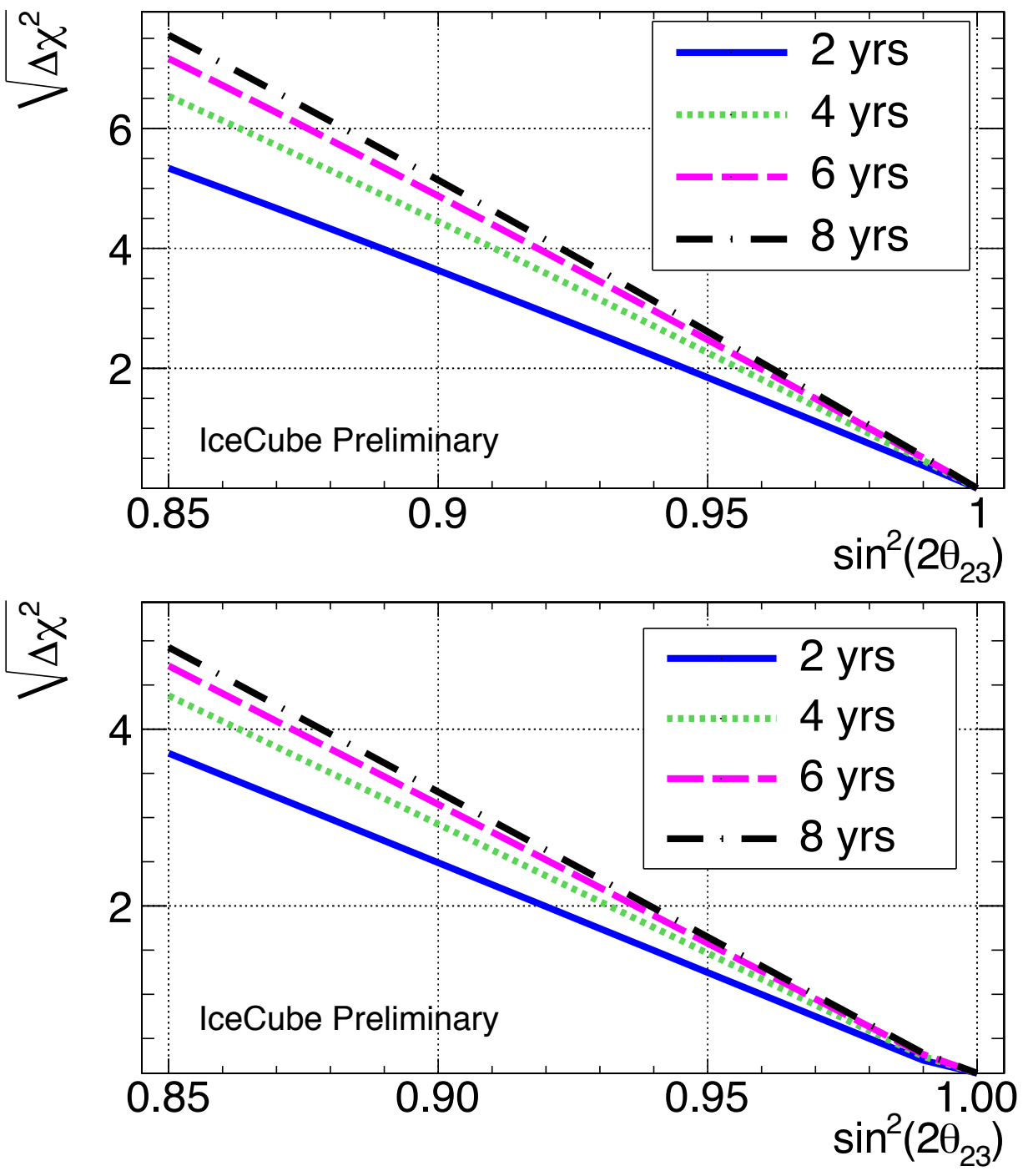

**Figure 1**: Sensitivity to the exclusion of maximal mixing without systematic uncertainties (top) and with systematic uncertainties (bottom) for various neutrino sample sizes for $\Delta m^2_{32} = 2.4 \cdot 10^{-3}\,\mathrm{eV}^2$. The graphs show the $\Delta\chi^2$ for maximal mixing compared to the best fit as a function of the injected mixing angle and minimized over $\Delta m^2_{32}$.

A better understanding of the experimental and theoretical systematic uncertainties will improve the sensitivity to the atmospheric oscillation parameters. While we are limited by statistics at the moment, a $5\sigma$ exclusion of non-maximal mixing with the given assumptions about background rejection and energy and zenith angle resolution will require reductions in our present systematic uncertainties. The systematic uncertainty on the atmospheric electron neutrino flux has the largest effect on the sensitivity, where it almost doubles the uncertainty on $\sin^2(2\theta_{23})$. This is mainly because the distribution of the electron neutrino events is similar to the distribution of the muon neutrino deficit. A similar effect is observed for tau neutrinos which cannot be distinguished from electron neutrinos. They appear at a rate proportionally to the muon neutrino deficit. An improved electron neutrino background rejection with a better particle identification will help to reduce the effect of the atmospheric electron neutrino flux uncertainty on the sensitivity. The most limiting experimental systematic uncertainty is the optical efficiency of the IceCube DOMs. This uncertainty will be reduced by calibration efforts in the lab and *in situ*.

Furthermore this study is based on the 79 string configuration, the following years of data are taken with the 86 string configuration containing two additional DeepCore strings with shorter string-to-string spacing. Additional DeepCore strings should improve the reconstruction quality and therefore also the background rejection. The shorter string distances will also help to lower the energy threshold.

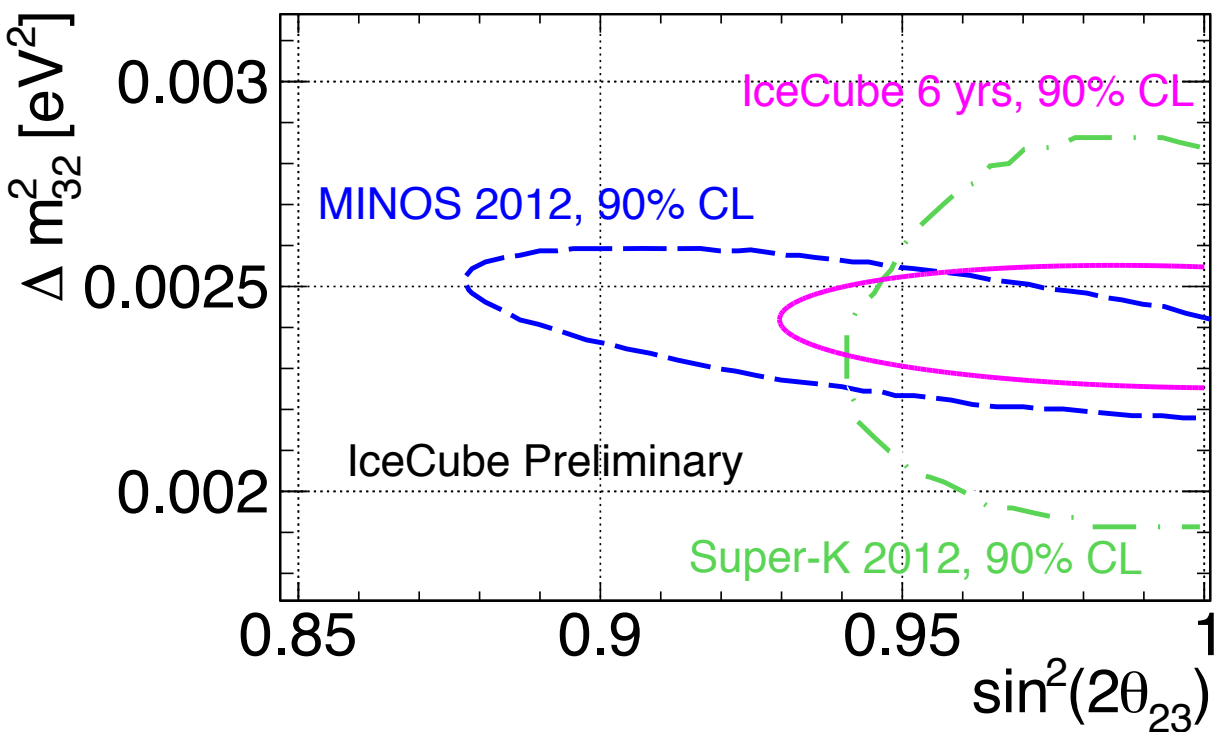

**Figure 2**: Confidence region with systematic uncertainties (90% CL) for six years of data for $\Delta m^2_{32} = 2.4 \cdot 10^{-3}\,\mathrm{eV}^2$ and $\sin^2(2\theta_{23}) = 1$ in comparison with recent results from MINOS [16] and Super-K [17].

# Measurement of atmospheric neutrino oscillations with IceCube/DeepCore in its 79-string configuration

THE ICECUBE COLLABORATION[1]

[1]*See special section in these proceedings*

*wiebusch@physik.rwth-aachen.de*

**Abstract:** With its low-energy extension DeepCore, the IceCube Neutrino Observatory at the Amundsen-Scott South Pole Station is able to detect neutrino events with energies as low as 10 GeV. This permits the investigation of flavor oscillations of atmospheric muon neutrinos by observing their zenith direction and energy. Maximum disappearance is expected for vertically upward moving muon neutrinos at around 25 GeV. A recent analysis has rejected the non-oscillation hypothesis with a significance of about 5 standard deviations based on data obtained with IceCube while it was operating in its 79-string configuration [1]. The analysis presented here uses data from the same detector configuration, but implements a more powerful approach for the event selection, which yields a dataset with significantly higher statistics of more than 8 000 events. We present new results based on a likelihood analysis of the two observables zenith angle and energy. The non-oscillation hypothesis is again rejected with a significance of about 5.1 standard deviations. In the 2-flavor approximation, our best-fit oscillation parameters are $\Delta m^2_{32} = (2.45 \pm 0.7) \cdot 10^{-3} \mathrm{eV}^2$ and $\sin^2(2\theta_{23}) = 1.0^{+0}_{-0.15}$.

**Corresponding authors:** Sebastian Euler[1] Laura Gladstone[2] Christopher Wiebusch[1]
[1] *III. Physikalisches Institut, RWTH Aachen University, D-52056 Aachen, Germany*
[2] *Dept. of Physics, University of Wisconcin, Madison, WI 53706, USA*

## 1 Introduction

Flavor oscillations of atmospheric neutrinos have been established by a wide range of experiments. Recently, they have also been observed by neutrino telescopes [1, 2], in an energy range above 10 GeV, previously not covered by other experiments. With the results presented here, IceCube improves the precision of the measurements in this energy range.

IceCube is a cubic-kilometer size neutrino detector installed in the ice at the geographic South Pole [3] between depths of 1450 m and 2450 m. Detector construction finished in December 2010, when all 86 strings of 60 detector modules each had been deployed. The event reconstruction relies on the optical detection of Cherenkov radiation emitted by secondary particles produced in neutrino interactions in the surrounding ice. Of fundamental importance for the analysis presented here is the DeepCore subarray. In the configuration that is used in this analysis, DeepCore consists of 6 densely instrumented strings plus 7 adjacent standard strings. It lowers IceCube's threshold to neutrino energies as low as 10 GeV, and thus opens a new window on the physics of atmospheric neutrino oscillations [4]. Atmospheric muon neutrinos moving vertically upwards through the detector have traveled roughly 12 700 km through the Earth since their production in the atmosphere of the northern hemisphere. For these events, maximum disappearance is expected around 25 GeV due to flavor oscillations. For smaller zenith angles the disappearance maximum shifts to lower energies. Fig. 1 shows the expected zenith angle and energy-dependent pattern in the muon neutrino survival probability. A previous IceCube analysis [1] has established the observation of neutrino oscillations and rejected the non-oscillation hypothesis with a significance of more than 5 standard deviations using standard methods for event selection and reconstruction. The analysis presented here uses data taken with the same detector configuration with 79 strings between May 2010 and May 2011, but implements different event selection and reconstruction techniques and aims for an improved measurement of the oscillation parameters $\Delta m^2_{32}$ and $sin^2(2\theta_{23})$. For our energy range, the simple 2-flavor formalism is adequate to describe neutrino oscillations. The survival probability in this formalism is given by

$$P(\nu_\mu \to \nu_\mu) = 1 - \sin^2(2\theta_{23}) \sin^2\left(1.27 \Delta m^2_{32} \frac{L}{E}\right)$$

with the neutrino propagation length $L$ in km and the neutrino energy $E$ in GeV.

## 2 Event selection

The primary background for this analysis is downward-going cosmic-ray-induced muons. Only a small number of these events are misreconstructed as coming up through the Earth, but the high rate would still dominate over the rate of the atmospheric neutrino signal without further event selection. Rejection of this background is usually achieved in IceCube analyses by requiring a high reconstruction quality, such as the goodness of the track fit. This approach yields a sample of high-quality events, but introduces selection biases in the observed distributions of zenith angle and energy. Higher energy events naturally are easier to reconstruct, leading to "better" values in the reconstruction quality variables, and are therefore preferred by an event selection relying on this type of variables. These potential biases must be carefully accounted for through simulation. For this analysis a different approach has been developed, which searches for starting events, which cannot be induced by atmospheric muons from cosmic-ray air showers and are therefore a clear signature of a neutrino interaction. This approach uses the outer layers of IceCube to reject atmospheric muons and achieves more unbiased distributions of zenith angle and energy.

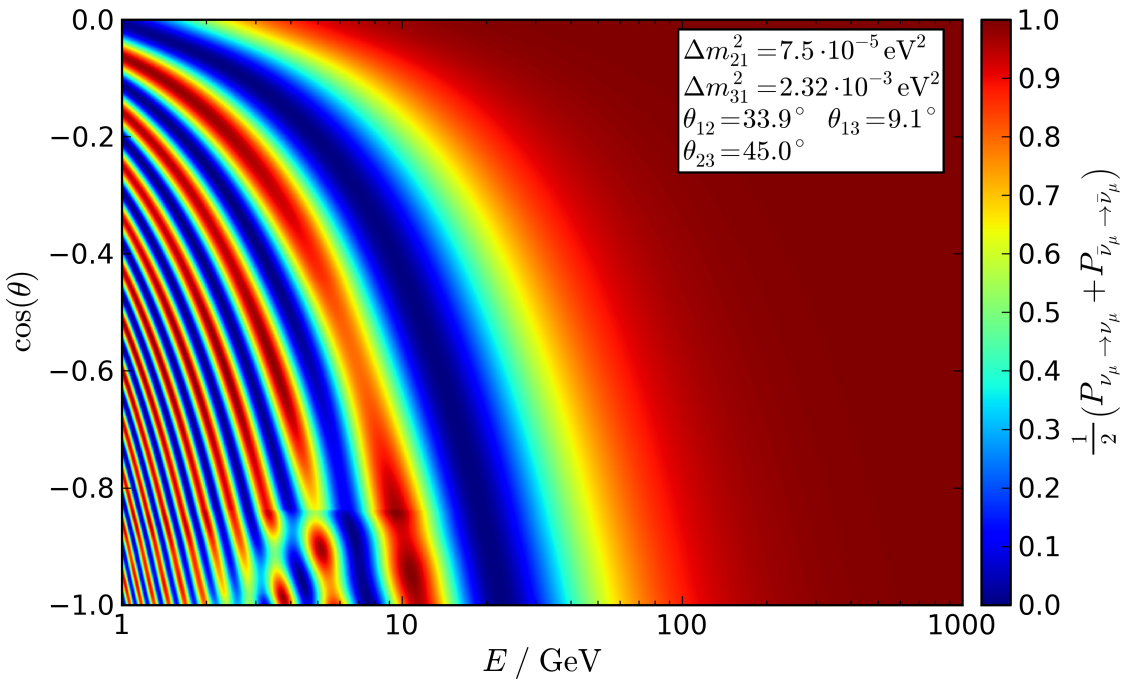

**Figure 1**: Survival probability of atmospheric muon neutrinos, depending on neutrino zenith angle and energy. Calculations were made with the tool nuCraft [5], which includes 3-flavor oscillations and matter effects. For energies $> 10$ GeV the 2-flavor approximation is adequate and thus used for this analysis.

Different veto techniques are employed throughout the event selection. First, an online filter algorithm rejects events based on their *particle speed* which is defined by the times of hits (i.e. detector modules with a signal) in the veto region relative to the center of gravity of hits in DeepCore [4]. This step reduces the background from misreconstructed cosmic-ray muons by more than an order of magnitude, while keeping more than 99% of the desired signal. Another example of a higher-level veto algorithm is illustrated in Fig. 2. First, we define as a reference the hit that fulfilled the DeepCore trigger condition. For all other hits we calculate the distance and time difference with respect to that reference hit. In the definition used, positive time differences are given by hits which occur before the trigger, negative time differences stem from later hits. In this projection, a particle entering the detector from the outside, triggering the detector, and then leaving the detector would move from top to bottom of the figure, approximately along the lines defined by the speed of light $c$. Thus, hits found along the line in the upper half are an indication for an incoming muon, whereas hits along the line in the lower half indicate a track leaving the detector. A simple way to identify background muons is to simply count the number of hits within an area along the "incoming muon" line of Fig. 2. In this analysis, events with more than 2 hits in the shaded area (the "veto hit region") are rejected.

The final event selection has been developed on simulated data. The background of atmospheric muons is simulated using the CORSIKA software [6]. Atmospheric neutrinos are simulated using the NuGen package [7] developed within the IceCube collaboration. The prediction by Honda et al. [8] is used to model the atmospheric neutrino spectrum. Note, that the cross sections implemented in NuGen do not reach below 10 GeV and include only deep inelastic scattering above. However, only an insignificant fraction of about 3% of our event sample has energies below

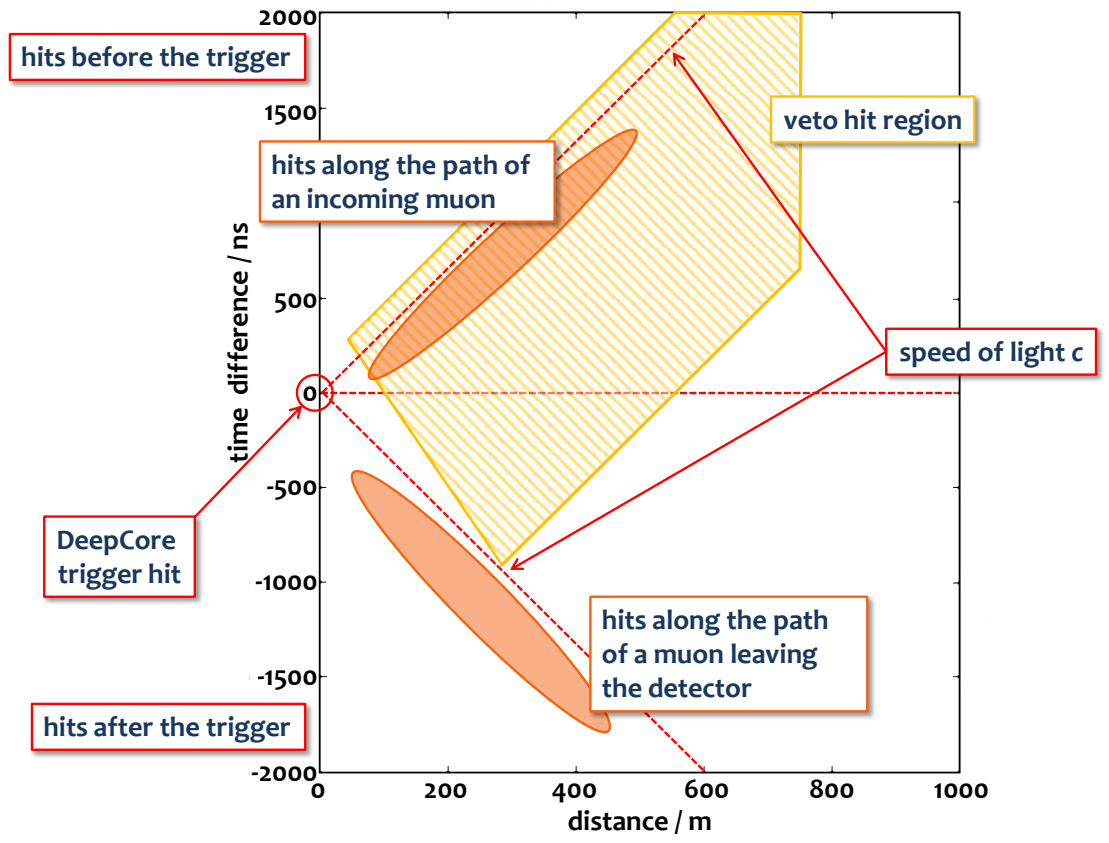

**Figure 2**: Illustration of the algorithm used for vetoing atmospheric muons. Events with more than 2 hits in the region along the "incoming muon" line are rejected.

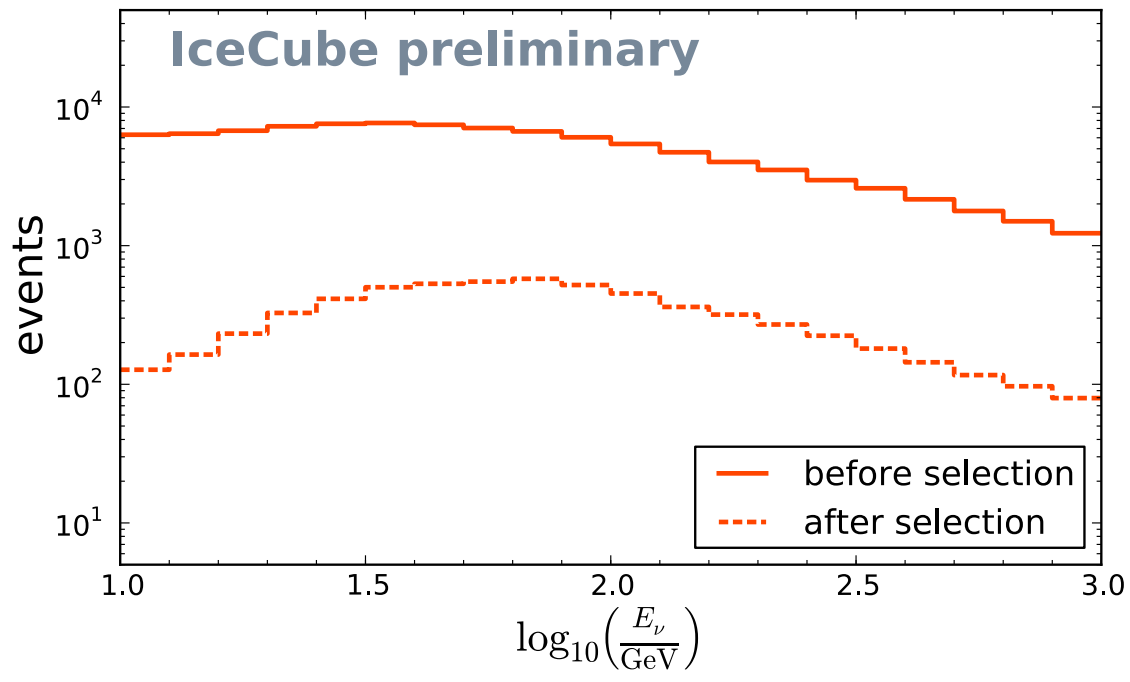

**Figure 3**: Energy spectrum of the simulated $\nu_\mu$ after the online filter (solid) and in the final event sample (dashed), assuming standard oscillation parameters $\Delta m^2_{32} = 2.4 \cdot 10^{-3}\mathrm{eV}^2$ and $\sin^2(2\theta_{23}) = 1.0$.

10 GeV. Other programs like GENIE [9] use more precise cross sections, but do not cover the whole higher energy range needed here. GENIE is, however, used for a more precise simulation of appearing $\nu_\tau$ events.

Other selection criteria include further veto cuts like the number of hits in the DeepCore region vs. the number of hits in the veto region, cuts evaluating the causality relation between hits (to reject noise-dominated events), and finally a selection of upward-going tracks with a reconstructed length of at least 40 m. To reject remaining background events, very soft cuts on selected reconstruction quality variables are applied.

The energy spectrum of the remaining muon neutrino events as expected from simulation (assuming standard oscillation parameters) is shown in Fig. 3. It peaks around 70 GeV and retains high statistics down to 10 GeV, throughout the energy range where oscillation effects are expected.

The remaining experimental data sample with a livetime of 312.3 days contains 8 117 events. The neutrino purity is estimated to be better than 90%, about 70% of which are expected to be muon neutrinos.

## 3 Reconstruction performance

In this analysis the oscillation parameters are derived from a comparison of reconstructed zenith angle and energy with the expectation from the simulation. Hence, the performance of the reconstruction is critical.

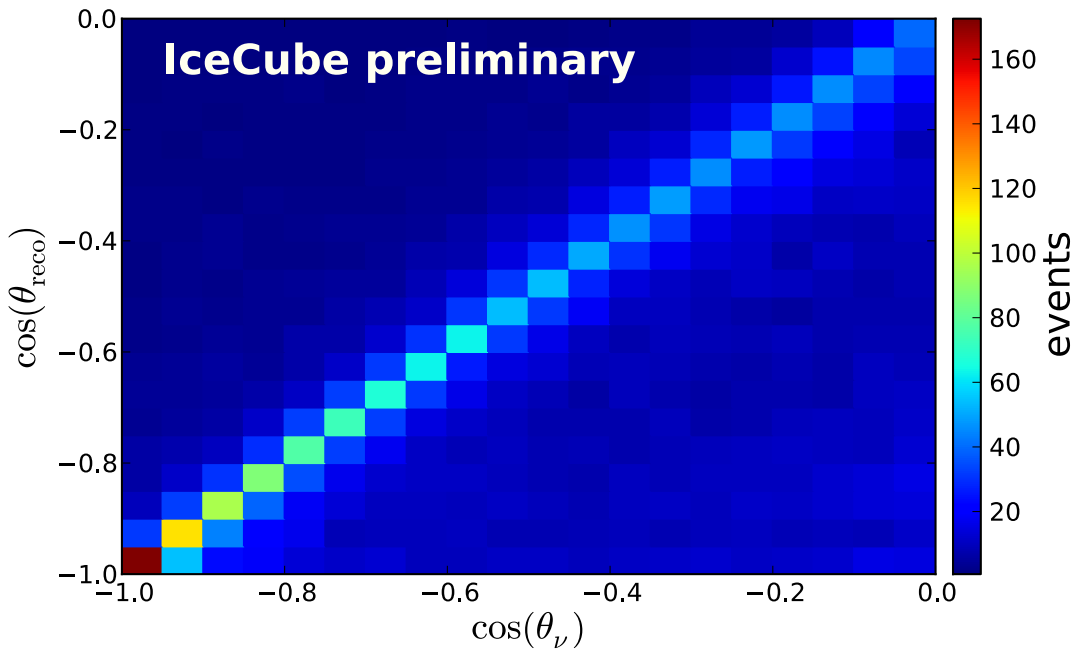

**Figure 4**: Correlation of reconstructed and true neutrino zenith angle for the selected $\nu_\mu$.

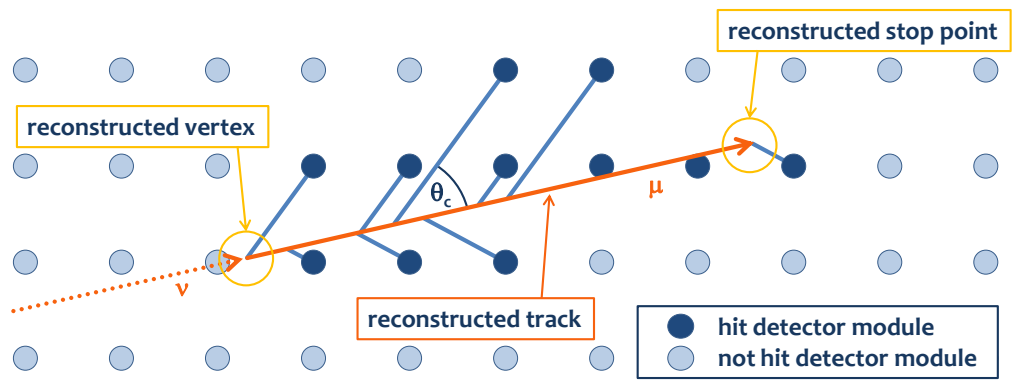

**Figure 5**: Illustration of the length reconstruction technique.

## 3.1 Zenith angle

Standard IceCube tools are used for the reconstruction of the zenith angle. As a first guess the *improved linefit* algorithm is used, followed by an iterative likelihood reconstruction (*MPEFit*) [10, 11]. The performance estimated from simulation of this final zenith angle reconstruction is shown in Fig. 4. In the relevant zenith angle region, the reconstruction achieves a median resolution of about 4°–8° for the $\nu_\mu$ sample.

## 3.2 Energy

As a proxy for the neutrino energy we use the reconstructed length of the muon track. In the energy range essential for this analysis, the muon track length is correlated to the neutrino energy: 1 GeV muon energy corresponds to about 4–5 m track length. The reconstruction agorithm *FiniteReco* [12] used estimates the track length by projecting all detected signals on the previously (by *MPEFit*) reconstructed track. The outermost projected points along the track define the reconstructed starting point (or *vertex*) and stopping point. The distance between these points is the reconstructed length. The principle is illustrated in Fig. 5.

The length reconstruction achieves a resolution of about 40 m. Fig. 6 shows the correlation between the reconstructed track length and the neutrino energy.

# 4 Analysis method

Simulated and measured data are binned into two-dimensional histograms of reconstructed zenith angle and track length. The binning covers 10 bins in cos(zenith angle) between $-1.0$ (vertical) and 0.0 (horizon) and 5 bins in log10(track length) between 1.5 and 3.0. For the simulation, separate histograms are made for each of the three neutrino flavors and for atmospheric muons. These four histograms can be separately weighted according to the disappearance and appearance probabilities for each flavor. They are then added to create a combined simulation prediction, representing a particular choice of oscillation parameters $\Delta m^2_{32}$ and $\sin^2(2\theta_{23})$ as well as normalizations of the different components. The combined simulation is fitted to the data by maximizing a global likelihood.

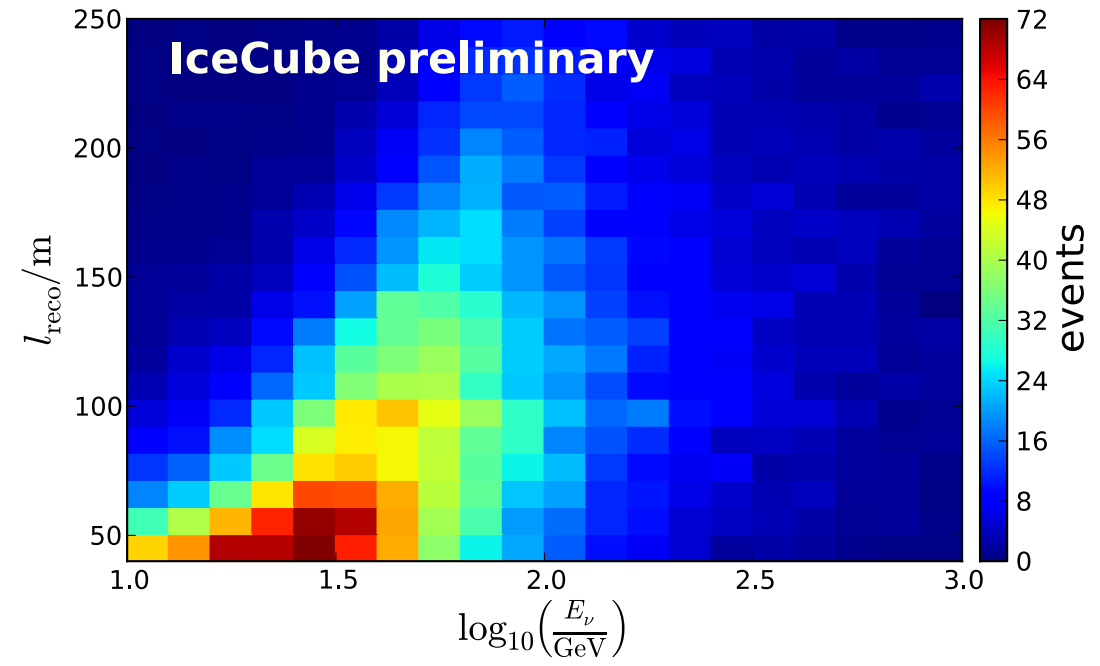

**Figure 6**: Correlation of reconstructed track length and neutrino energy for the selected $\nu_\mu$.

For the likelihood, we use the standard Poisson formulation, where we calculate for each bin $(i,j)$ the probability to observe $d_{ij}$ events in the measured data, given $s_{ij}$ events in the simulated data. In addition to fitting the oscillation parameters, five nuisance parameters are also left free in the fit: the normalizations of the individual simulation components (a common normalization for $\nu_\mu$ and $\nu_\tau$, and separate normalizations for $\nu_e$ and atm. $\mu$), the spectral index $\gamma$ of the primary cosmic ray spectrum and the relative contribution of pions and kaons to the neutrino flux. These nuisance parameters absorb certain systematic uncertainties in the primary cosmic ray flux, the $\nu_\mu/\nu_e$ production ratio, uncertainties in the neutrino cross sections and the overall optical efficiency of the detector. Note, that keeping a broad range in energy and zenith angle – also of regions unaffected by oscillations – is important for constraining the absolute normalizations and other nuisance parameters in the fit. Our knowledge of systematic uncertainties is reflected by Gaussian priors for each nuisance parameter $k$, which are added to the likelihood. Table 1 gives their central values and uncertainties. The full likelihood expression has the form

$$-LLH = \sum_{i,j}(s_{ij} - d_{ij}\ln(s_{ij})) + \frac{1}{2}\sum_k \left(\frac{q_k - \langle q_k\rangle}{\sigma_k}\right)^2 .$$

Other systematics, which are not directly implemented in the fit, are evaluated by separate simulations. These include the optical efficiency and variations in the description of the ice properties of the bulk of the detection medium as well as of the refrozen ice around the strings.

| Nuisance parameter | $\langle q_k\rangle$ | $\sigma_k$ |
|---|---|---|
| $\nu_\mu$ & $\nu_\tau$ norm. | 1.0 | 25% |
| $\nu_e$ norm. | $\nu_\mu$ & $\nu_\tau$ norm. | 20% |
| atm. $\mu$ norm. | — no constraint — | |
| spectral index $\gamma$ | 2.65 | 0.05 |
| $\pi$/K ratio | 1.0 | 10% |

**Table 1**: Central values and uncertainties of the Gaussian priors for the nuisance parameters.

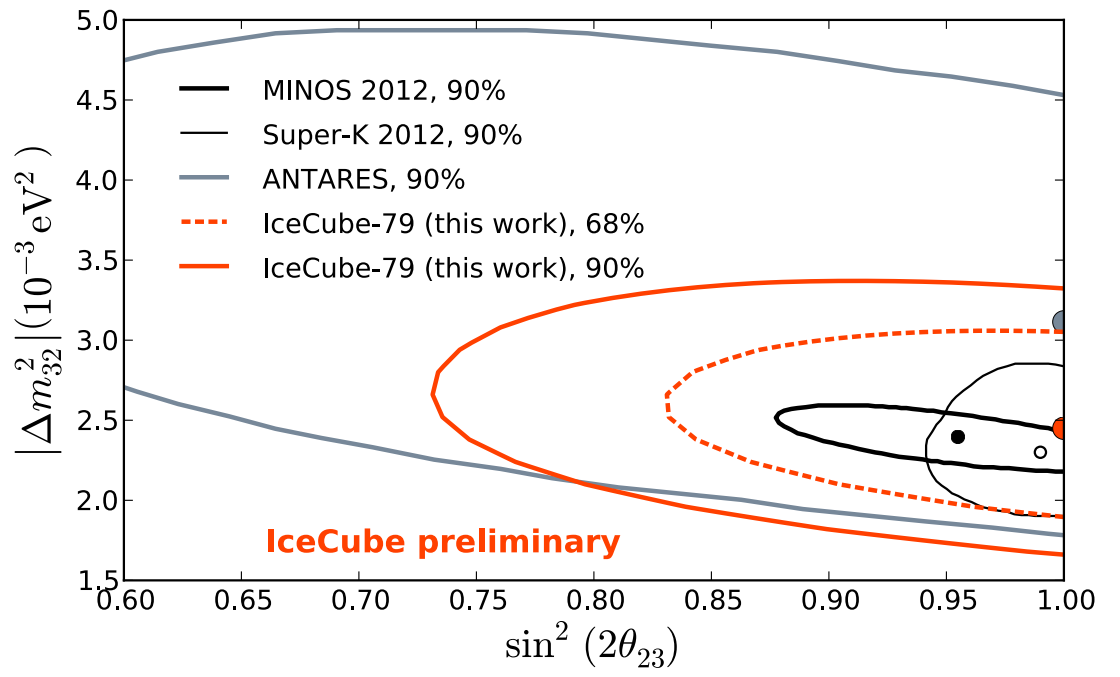

**Figure 7**: Confidence regions of the fitted oscillation parameters, together with recent results from MINOS [14], Super-Kamiokande [15], and ANTARES [2].

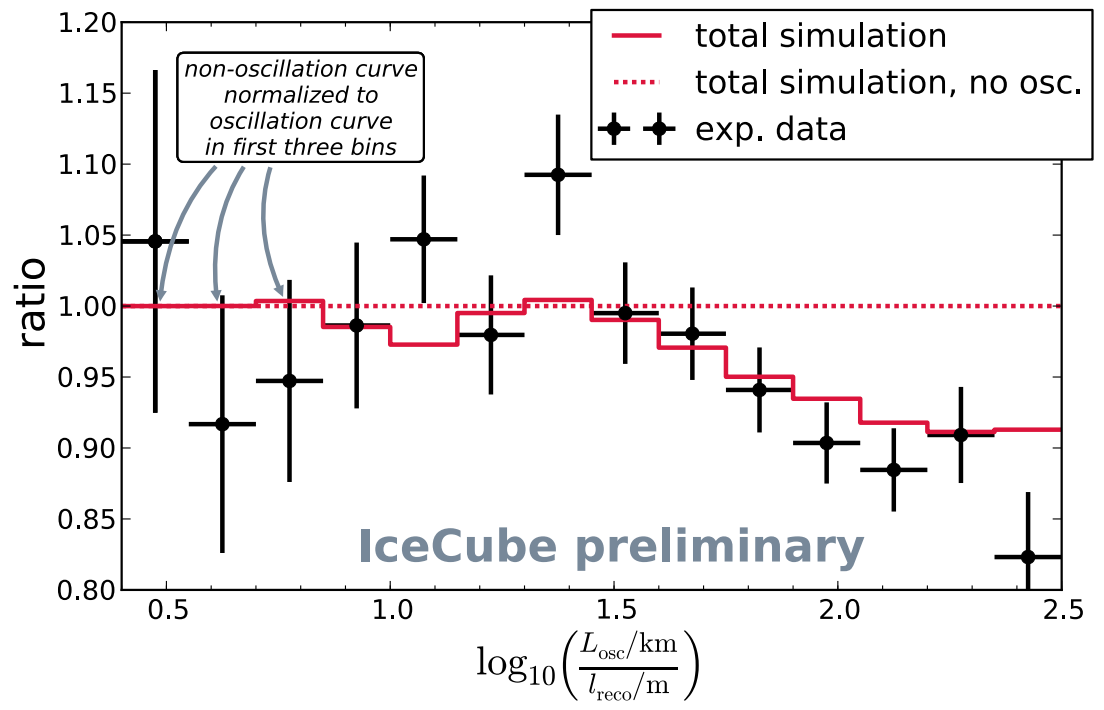

**Figure 8**: Distribution of oscillation length divided by reconstructed track length, ratio of experimental data and best-fit simulation to the non-oscillation hypothesis.

## 5 Results

A scan of the oscillation parameter space was performed. First, the best-fit oscillation (and nuisance) parameters were determined by maximizing the global likelihood, as described above. Then, for each point in the oscillation parameter space, a minimization of only the nuisance parameters was done. The ratio of the likelihood of the fit at each point to the fitted global maximum is used to calculate the regions in the oscillation parameter space that are compatible with our observations. Preliminary significances in units of standard deviations are calculated according to Wilks' theorem [13]. Fig. 7 shows the resulting 68% and 90% confidence regions, together with the best-fit point.

The measured effect of neutrino oscillations can be visualized by plotting the ratio of oscillation length and neutrino energy. For this analysis, this translates to the ratio of oscillation length, calculated from the measured zenith angle, and reconstructed track length, which serves as our energy proxy. Fig. 8 shows this quantity for data and best-fit simulation, relative to the best-fit non-oscillation hypothesis, which has been normalized to the oscillation curve in the first three bins, where no oscillation effects are expected.

As a preliminary result of this analysis, the non-oscillation hypothesis is rejected by a likelihood ratio corresponding to 5.1 standard deviations. As best-fit oscillation parameters we find $\Delta m^2_{32} = (2.45 \pm 0.7) \cdot 10^{-3}\mathrm{eV}^2$ and $\sin^2(2\theta_{23}) = 1.0^{+0}_{-0.15}$. Fig. 9 shows track length and zenith angle distributions for the best-fit oscillation and nuisance parameters.

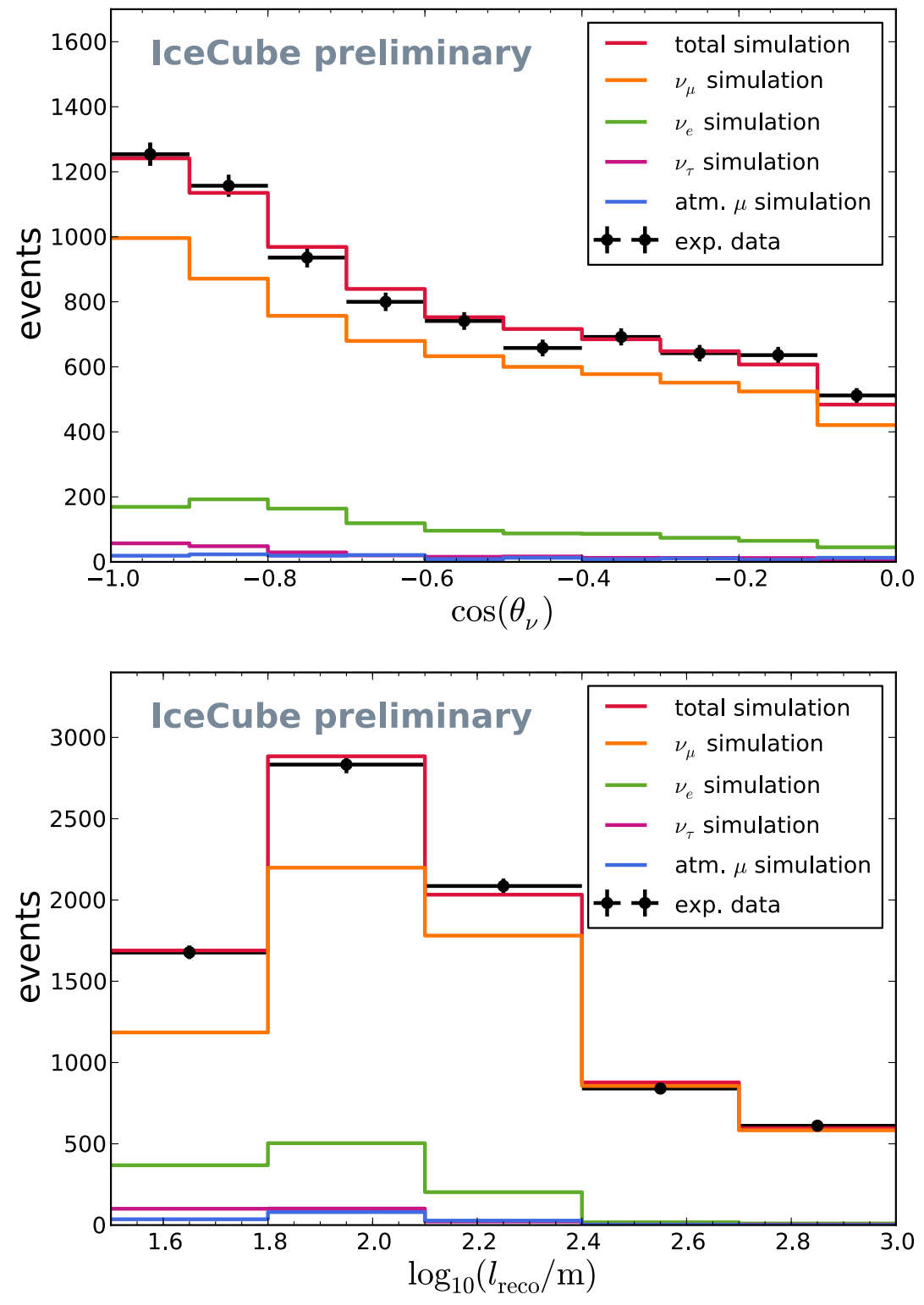

**Figure 9**: Distributions of reconstructed zenith angle and track length, with best-fit oscillation and nuisance parameters.

Future analyses might profit from significantly improved reconstruction techniques (see e.g. [16]), from further refinements of the event selection, and from the even higher statistics of multi-year datasets. The techniques established in this analysis are expected to eventually qualify IceCube to deliver a competitive measurement of the oscillation parameters (see [17]).